\documentclass[10pt]{article}

\usepackage{graphicx}
\usepackage{amsmath,amssymb}
\usepackage{paralist, tabularx}
\usepackage{dcolumn}
\usepackage{ifsym}
\usepackage{bm}
\usepackage{textgreek}
\newcommand{\eps}{\varepsilon}
\newcommand{\nc}{{n_\mathrm{c}}}
\newcommand{\mc}{m_\mathrm{c}}
\newcommand{\muc}{\mu_\mathrm{c}}
\newcommand{\Veq}{V_\mathrm{eq}}
\newcommand{\Ivc}{I_\mathrm{vc}}
\newcommand{\Ivcsm}{I_\mathrm{vc,sm}}
\newcommand{\dotIvcsm}{\Dot{I}_\mathrm{vc,sm}}
\newcommand{\etavc}{\eta_\mathrm{vc}}
\newcommand{\Vcc}{V_\mathrm{cc}}
\newcommand{\Ieq}{I_\mathrm{eq}}
\newcommand{\Iext}{I_\mathrm{ext}}
\newcommand{\yeq}{y_\mathrm{eq}}
\newcommand{\yl}{y_\mathrm{ad}}
\newcommand{\yr}{y_\mathrm{ev}}

\newcommand{\afix}{\operatorname{alg}_{\operatorname{fix}}}
\newcommand{\tran}{\mathsf{T}}
\newcommand{\rev}[1]{\textcolor{black}{#1}}
\def\gc{g_{\mathrm{c}}}
\def\Vh{V_{\mathrm{h}}}
\def\Ih{I_{\mathrm{h}}}

\def\dotVh{\dot{V}_{\mathrm{h}}}
\def\dotIh{\dot{I}_{\mathrm{h}}}

\def\gc{g_{\mathrm{c}}}
\def\Vh{V_{\mathrm{h}}}
\def\Ih{I_{\mathrm{h}}}

\def\dotVh{\dot{V}_{\mathrm{h}}}

\DeclareMathOperator{\diag}{diag}

\DeclareMathOperator{\bias}{bias}
\usepackage{xcolor,hyperref}
\definecolor{darkblue}{rgb}{0.0,0.0,0.5}
\definecolor{darkgrey}{rgb}{0.5,0.1,0.1}
\hypersetup{colorlinks,breaklinks,
            linkcolor=darkblue,urlcolor=darkblue,
            anchorcolor=darkblue,citecolor=darkblue}

\title{\textbf{Observing hidden neuronal states in experiments}}
\author{Dmitry Amakhin\textsuperscript{1}, Anton Chizhov\textsuperscript{2}, Guillaume Girier\textsuperscript{3},\\ Mathieu Desroches\textsuperscript{4,*}, Jan Sieber\textsuperscript{5,*}, Serafim Rodrigues\textsuperscript{3,6,*}}
\date{}

\begin{document}
\maketitle
\begin{center}
\textsuperscript{1}{Laboratory of Molecular Mechanisms of Neural Interactions,\\ Sechenov Institute of Evolutionary Physiology and Biochemistry of RAS, Saint Petersburg, Russia}\\
\textsuperscript{2}{Institute for Theoretical Physics, University of Bremen, Bremen, Germany}\\
\textsuperscript{3}{MCEN Team, Basque Center for Applied Mathematics (BCAM), Bilbao, Spain}\\
\textsuperscript{4}{MathNeuro Project-Team, Inria Branch of the University of Montpellier, Montpellier, France}\\
\textsuperscript{5}{College of Engineering, Mathematics and Physical Sciences, University of Exeter, Exeter, United Kingdom}\\
\textsuperscript{6}{Ikerbasque, the Basque Foundation for Science, Bilbao, Spain}

\noindent * mathieu.desroches@inria.fr; Jan.Sieber@exeter.ac.uk; srodrigues@bcamath.org
\end{center}

\begin{abstract}
In this article we demonstrate a general protocol for constructing systematically experimental steady-state bifurcation diagrams for electrophysiologically active cells. We perform our experiments on entorhinal cortex neurons, both excitatory (pyramidal neurons) and inhibitiory (interneurons). \rev{A slowly ramped voltage-clamp electrophysiology protocol serves as closed-loop feedback controlled experiment for the subsequent current-clamp open-loop protocol on the same cell. In this way, the voltage-clamped experiment determines dynamically stable and unstable (hidden) steady states of the current-clamp experiment.
The transitions between observable steady states and observable spiking states in the current-clamp experiment provide partial evidence for stability and bifurcations of the steady states. This technique for completing steady-state bifurcation diagrams in a model-independent way expands support for model validation to otherwise inaccessible regions of the phase space.} Overlaying the voltage-clamp and current-clamp protocols leads to an experimental validation of the classical \textit{slow-fast dissection} method introduced by J. Rinzel in the 1980s and routinely applied ever since in order to analyse slow-fast neuronal models. Our approach opens doors to observing further complex hidden states with more advanced control strategies, allowing to control real cells beyond pharmacological manipulations.
\end{abstract}

\section{Introduction}
When characterising the dynamics of nonlinear systems, one fundamental criterion for a model is if its stable states such as stationary solutions or periodic orbits match experimental observations. The ability to \rev{fit and} validate models is, thus, greatly expanded by experimental tools with the capacity to unveil non-observable (sensitive or dynamically unstable) states that are otherwise inaccessible to standard measurements. The combination of observable and non-observable states gives access to an experimental equivalent of parameter-dependent families of stable and unstable states in a model, which are usually referred to as a bifurcation diagram. 

This article applies a new experimental technique of using feedback control to find unstable states to electrophysiology experiments on neuronal cells. Our aim is to support systematic validation of neuron models by comparing bifurcation diagrams and observing their between-cell variability. We focus on unstable parts of input-dependent families of steady-state solutions obtained by feedback-controlled experiments and compare them with indirect evidence from standard measurements from open-loop experiments. This extends recent work of Ori \emph{et al.}~\cite{ori2018,ori2020} constructing phase diagrams from neuronal data, and complements other approaches such as using data to verify the bifurcation structure of neuronal models~\cite{levenstein2019,hesse2022}, model-based data analysis~\cite{sip2023} or parameter estimation from data~\cite{ladenbauer2019}.

Our technique is a simplification of the so-called \textit{control-based continuation (CBC)} method, an approach which has been recently demonstrated in mechanical experiments~\cite{sieber2008a}, vibrations and buckling experiments~\cite{RENSON2019449}, pedestrian flow experiments~\cite{panagiotopoulos2022control}, atomic-force microscopy \cite{bottcher2025exposing}, cylindrical pipe flow simulations~\cite{willis_duguet_omelchenko_wolfrum_2017}, and feasibility studies for synthetic gene networks~\cite{de2022control,blyth2023numerical}. Indeed, CBC is a procedure that combines feedback control~\cite{phillips1999} and pseudo-arclength continuation~\cite{allgower2003} in a model-free environment, that is, only reliant upon noisy experimental data in a closed-loop control setup. The objective is to compute experimental bifurcation diagrams, that is, both families of stable and unstable states (either stationary or periodic) together with bifurcation points joining such families. Importantly, the control is noninvasive in the sense that it vanishes once an equilibrium or periodic of the uncontrolled system has been reached; the control is decreased iteratively using a Newton's method. In our case, the approach is simpler and does not require the use of Newton iterations since we consider a slow ramp on the control target. Hence we exploit de presence of two timescales in the experimental procedure in order to obtain directly an approximation of the steady-state bifurcation curve, and we explain why such a bifurcation curve is obtained at low cost; see below. \rev{Our approach is hence useful for both modelers and experimentalists, as it relies upon standard protocols routinely used in patch-clamp electrophysiology. It can further decipher the excitability class of a given real neuron as well as help fitting a model to data by using both current-clamp and voltage-clamp protocols.}

\begin{figure*}[!t]
\includegraphics[width=\textwidth]{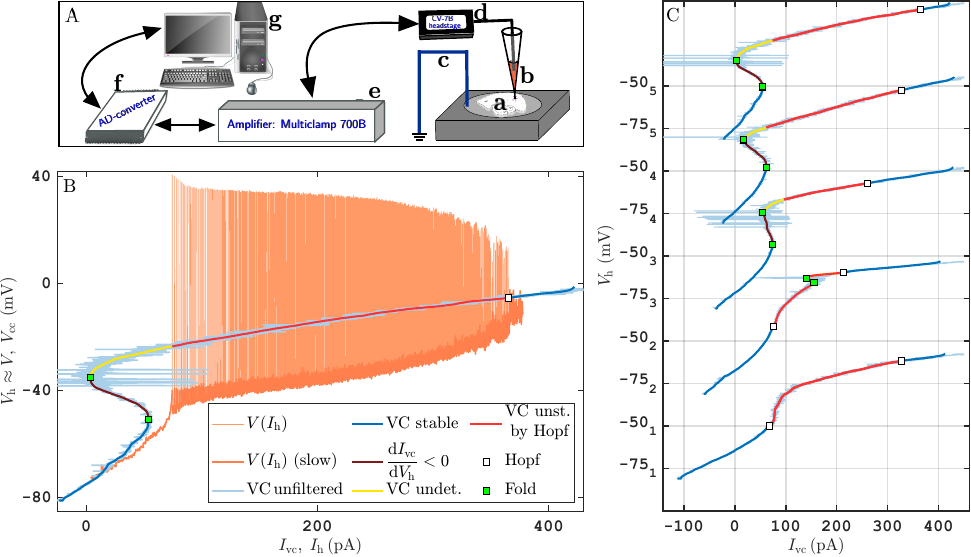}
\caption{A: {Sketch of experimental setup} with brain slice ({\textbf{a}}),  patch pipette ({\textbf{b}}), reference electrode (Ag-AgCl pellet) connected to ground ({\textbf{c}}),  amplifier ({\textbf{e}}, \textit{Multiclamp 700B}) with CV-7B headstage ({\textbf{d}}), AD-converter ({\textbf{f}}, \textit{National Instruments NI USB-6343}) and standard PC computer ({\textbf{g}}, \href{https://pixabay.com/vectors/computer-desktop-workstation-office-158675/}{https://pixabay.com/vectors/computer-desktop-workstation-office-158675/}).
B: {VC and CC protocol runs for cell $5$}: $(\Ivc(t),\Vh(t))$-curve for VC run (thin bright blue: unfiltered data with sampling time step $5\times10^{-5}$\,s, blue/red/brown: median of $\Ivc$ over moving windows of size $\Delta w=4\times10^{3}$ steps equalling $0.2$s) and $(\Ih(t),\Vcc(t))$-curve for CC run (orange, mean of $\Ih$ over moving windows. C: $(\Ivc(t),\Vh(t))$-curves for VC protocol of all $5$ cells on waterfall $\Vh$-axis (color coding indicates conjectured stability as indicated for (b)).
\label{fig:fig1}}
\end{figure*}

In our electrophysiology experiments on entorhinal cortex neurons we treat the cell as an electric circuit, and apply a voltage clamp (VC)~\cite{cole1955,hodgkin1952b}, followed by current clamp (CC). In the VC setup the electrode acts as a voltage source at the neural membrane, fixing the potential across the neural membrane, measuring the current, while the CC setup adds a fixed external current, measuring the resulting membrane potential; see the illustration for the experimental setup in Fig~\ref{fig:fig1}A and the S1 Text for further details. In-vivo neurons are subject to current signals  that drive spiking (oscillatory) or rest (steady) states of the neural membrane potential~\cite{schaefer2006}. This makes the CC setup the open-loop part of the protocol. In contrast, the VC part of our experiment is the closed-loop feedback-controlled part of the protocol because the voltage source regulates the external current to maintain the hold voltage. VC has been applied successfully  to study neuronal nonlinear current-voltage relationships, so-called \textit{N-shaped $I$-$V$ characteristics}, which cause enhanced neuronal excitability and influence the regenerative activation of certain ionic currents (e.g., sodium)~\cite{schwindt1981,fishman1969,johnston1980,blatt1988,vervaeke2006} across the neural membrane, into and out of the cell. We show that the VC protocol with a slowly varied reference voltage signal gives access to stable and unstable neuronal steady states of the neuron, which was hinted at in~\cite{delnegro1998,qian2014}. In contrast, the open loop CC protocol with a slowly varying applied current always follows stable (observable) states, driving the neuron to dynamically transition between its observable rest states and its observable spiking states. 

We interpret these combined experimental protocols (VC and CC) using multiple-timescale dynamics, in particular, the dissection method~\cite{rinzel87}, which reveals the dynamic bifurcations in the experiment, as demonstrated in the experimental bifurcation diagram in Fig~\ref{fig:fig1}B. 
In this slow-fast framework, the states traced by the VC protocol with slow variation correspond to the steady-state experimental bifurcation diagram of the so-called \textit{fast subsystem} of the mathematical model describing the protocol~\cite{rinzel87}, that is, the model system with constant input current. Hence, the N-shaped I-V relation of a neuron should be seen as a S-shaped V-I bifurcation diagram, such as in Fig~\ref{fig:fig1}B.

Following this strategy, we demonstrate the feasibility of tracking a family of neuronal steady states (stable and unstable) via variations of reference signal and reparameterizing the obtained curve using the feedback current. 

\section{Results} 
The experimental neuronal bifurcation diagram in Fig~\ref{fig:fig1}B shows the time profiles $(\Ih(t),\Vcc(t))$ of the CC protocol run (orange, thin) and $(\Ivc(t),\Vh(t))$ of the VC protocol run (bright blue, thin) for cell $5$ overlaid in the $(I,V)$-plane. \rev{For both runs $\Ih(t)$ and $\Vh(t)$ were slowly increased, respectively (see Materials and Methods).} After smoothing, the VC time profile is the S-shaped curve $(\Ivcsm(\Vh),\Vh)$ (blue/brown/red, thick). It  equals the $(I,V)$-characteristic of the stationary neuronal states of the CC protocol, \emph{including dynamically unstable states} (brown and parts of red). The transition to stable spiking states is compatible with \rev{(Hodgkin) class}-I excitability, however we do not have sufficient data to distinguish the different classes of excitability. See Fig\,B in S1 Text for how the interspike intervals depend on $\Ih$.

The dynamical stability and instability of stationary states is inferred based on two pieces of evidence: (i) the negative slope of the $(\Ivcsm,\Vh)$-curve (brown) after smoothing over a moving window with larger size ($\Delta w=1.5\times 10^4$ steps equaling $=0.75$s in Fig~\ref{fig:fig1}C), or (ii) by the presence of slow-fast oscillations (neuronal firing) in the CC run at $\Ih$ equalling the $\Ivcsm$ (red).
Criterion (i) implies instability for topological reasons. \rev{Criterion (ii) provides only circumstantial evidence.
In Fig~\ref{fig:fig1}B (cell 5) the darker shading of the (orange-colored) CC run indicates slow dynamics as identified by the norm $|(\Vcc'(t)\sqrt{\Delta_t},\Vcc''(t))\Delta_t|$ being less than a fixed threshold (after smoothing of $\Vcc(t)$ and $\Vcc'(t)$, see S1 Text for detailed definition). This shading shows that the firing oscillations in Fig~\ref{fig:fig1}B spend most time near their voltage minimum $V_{\min}$ for the respective $\Ih(t)$. Fig~\ref{fig:fig5}C and \ref{fig:fig5}E show an embedding of the firing oscillations of cell 5 into the $(\Vcc,\Vcc')$-plane, and a zoom near $\Ih\approx 200$\,pA. The projection and zoom indicate that the firing oscillation passes slowly near the presumed fixed point, but leaves its neighborhood again. At the low-current end of the $(\Ivcsm,\Vh)$-curve, the CC protocol  for cell 5 shows no firing oscillations (recall that $\Ih$ is ramped up) in Fig~\ref{fig:fig1}B. Thus, the part of the $(\Ivcsm,\Vh)$ curve between fold and presence of slow-fast oscillations is labelled ``undetermined'' (colored yellow) as the combination of single-ramp VC and CC protocol do not provide evidence for or against stability of this part. We do not label the ``transition'' from stability label ``undetermined'' to ''unstable'' as a Hopf bifurcation as the slow parts of the firing oscillations are approximately $20$\,mV below the equilibrium indicated by the VC run. So, it is unclear if (and where precisely) a change of stability occurs between $I_\mathrm{db}\approx365$\,pA and the fold of steady states at $\Ih\approx3$\,pA.}

The stability boundaries of the stationary states are labelled as bifurcations in Fig~\ref{fig:fig1}B. The change of stability near the disappearance of stable spiking states at $I_\mathrm{db}$ is labelled as a Hopf bifurcation. \rev{We observe that for $\Ih>I_\mathrm{db}$ small-amplitude oscillations are visible, which emerge from relaxation type oscillations with a slow phase near the stationary state for $\Ih<I_\mathrm{db}$. Fig\,I in S1 Text shows a zoom of Fig~\ref{fig:fig1}B near $I_\mathrm{db}$. These features are typical for a singular Hopf bifurcation \cite{baer1986singular} as encountered also in model simulations for excitatory neurons. The fold points of the $(\Ivcsm,\Vh)$-curve are saddle-node (fold) bifurcations.} Fig~\ref{fig:fig1}C shows the stationary-state curves with their inferred dynamical stability for all $5$ cells (see S1 Text for CC run time profiles used to partially infer stability of cells $1$--$4$). The cells are vertically ordered and numbered according to depth of the S-shape, determining if they fall into the \rev{category of class-I or class-II} neurons. \rev{We use the same convention as in Fig~\ref{fig:fig1}B for parts of the curve where we have partial evidence for instability: no bifurcation is indicated at transitions between stability labels ``undetermined'' and ``unstable''.}
Fig~\ref{fig:fig1}C demonstrates wide variability in steady-state curve shape among cells of nominally same function.

\begin{figure*}[!t]
\centering
\includegraphics[width=\textwidth]{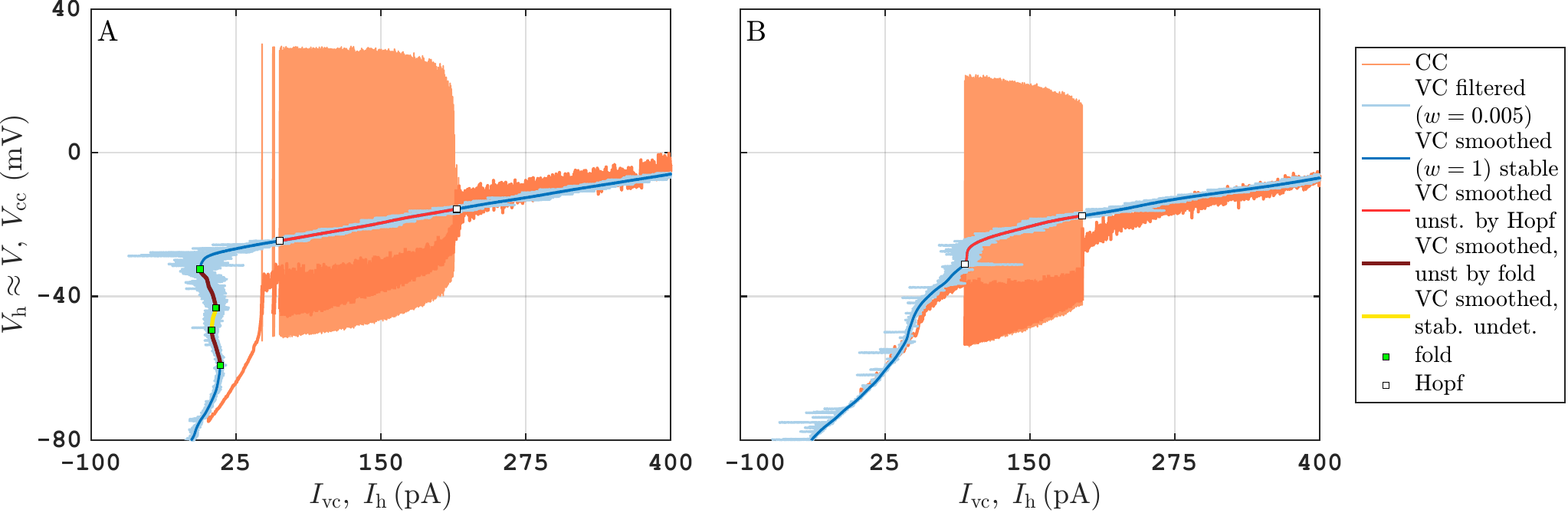}
\caption{A: Experimental bifurcation diagram for an interneuron from the entorhinal cortex. B: Experimental bifurcation diagram for a \rev{class}-II PY neuron from the same region. Protocols identical and color coding to Fig~\ref{fig:fig1}B.}
\label{fig:fig2}
\end{figure*}
The method described in this work is versatile and uses only standard electrophysiological procotols. Hence, it can be applied to any neuron type. The experimental bifurcation diagrams in Fig~\ref{fig:fig2}A illustrate this point. Panel A shows the results of the VC and CC protocols applied to an inhibitory neuron from the entorhinal cortex while panel B shows the results of same protocols applied to a \rev{class}-II neuron. The steady-state curve in Fig~\ref{fig:fig2}A is more complicated than for PY neurons' recordings shown in other figures. However it is compatible with a \rev{class}-I excitable neuron. \rev{Since the steady-state curve in Fig~\ref{fig:fig2}A has multiple folds, there are some parts of the branch where we cannot determine stability safely from topology (that is, by checking that $\mathrm{d}\Ivcsm/\mathrm{d}\Vh<0$), or from the near-by embedded trajectories of relaxation oscillations. Similar to Fig~\ref{fig:fig1} we have colored the parts of the steady-state curve in Fig~\ref{fig:fig2}A with undetermined stability yellow, while keeping the part for which instability is suggested by existence of relaxation oscillations in red.  For the range of $\Ih$ between $43$\,pA and $60$\,pA during the CC protocol in Fig~\ref{fig:fig2}A we observe a stable steady state $\Vcc$ response, with intermittent voltage spikes typical for an excitable stable steady state near a singular Hopf bifurcation subject to noise \cite{Lindner2004}. For this reason we label the low-$\Ih$ onset of large-amplitude relaxation oscillations in Fig~\ref{fig:fig2}A also as a Hopf bifurcation. A potential alternative mechanism for the sudden transition of firing oscillations to steady state near $I_\mathrm{db}$ ($\approx210$\,pA for Fig~\ref{fig:fig2}A, $\approx190$\,pA for Fig~\ref{fig:fig2}B) is that the firing oscillations experience a \emph{saddle-node of periodic orbits}. This would imply a bistability between firing and steady state for $\Ih<I_\mathrm{db}$, a scenario that has been observed near depolarization blocks in models of dopaminergic neurons by \cite{dovzhenok2012exploring}. Our illustrative examples of mathematical neuron models  shown in Figs~\ref{fig:fig3}A and \ref{fig:fig4} have such a small region of bistability (Fig~\ref{fig:fig3}A near $I_\mathrm{db}$ and Fig~\ref{fig:fig4} near $\Ih\approx135$\,pA) for the parameters used in the computations (see Figs\,F, G and H in S1 Text for numerical bifurcation diagrams).}

We observe that the steady-state curves for the CC run and the VC run slightly deviate from each other\rev{, which we attribute to a drift in cell properties}. There are examples of scenarios with multiple folds in the steady-state curve in the literature, in particular with low-threshold spiking neurons~\cite{rush1994}.

\subsection{Analysis}
To see why VC-run time profiles approximate the experimental bifurcation diagram with unstable stationary states of neurons in CC protocol, we use the formalism of multiple-timescale dynamical systems. Superimposing the data from VC and CC protocol also gives the first experimental illustration of slow-fast dissection. The effect of the respective clamps can be understood in a general class of conductance-based models for the neuron,
\begin{equation}\label{eq:gencond}
  \begin{aligned}
C\dot{V} &=-\textstyle{\sum}_jI_j(x_j,V)+\Iext,\\ 
\tau_j(V)\dot{x}_j&=x_{\infty,j}(V)-x_j,
\end{aligned}
\end{equation}
describing the current balance across the neuron's membrane. The
membrane potential is $V$, $I_j(x_j,V)_{j=1,\ldots,N}$ are the
currents across different voltage-gated ion channel types and $\Iext$
is the external current. Each channel type $j$ has a set of
  associated gating variables $x_j$, which are possibly vectors of
  length $n_j$ if the channel gate has both activation and
  inactivation states, with steady-state gating functions
$x_{\infty,j}(V)$ (also of size $n_j$) and
relaxation times $\tau_j(V)$ (a diagonal $n_j\times n_j$ matrix). The observed dynamic effects such as oscillations
(firing/spiking) and negative-slope $(I,V)$ characteristics are
determined by these channel coefficients $I_j$, $x_{\infty,j}$ and
$\tau_j$ that are traditionally obtained by parameter fitting from VC
experiments, a difficult and ill-posed problem, as gating
  variables $x_j$ are not directly measured~\cite{de2018}.

The VC and CC protocols use different mechanisms for generating $\Iext(t)$. The VC protocol is a closed loop where a voltage source regulates $\Iext$ with high gain $\gc$ to achieve the slowly varying hold voltage $\Vh$ at the voltage source for general model \eqref{eq:gencond}, measuring $\Ivc$: 
\begin{align}\label{eq:VCSR}
\Iext&\approx\Ivc=\gc(\Vh-V),& \dotVh&= \eps \Delta_V
\end{align}
which turns general model \eqref{eq:gencond} with VC protocol \eqref{eq:VCSR} into a multiple-timescale dynamical system with $1+\sum n_j$ fast state variables $(V,x_j)$ (where $\sum n_j$ is the overall number of gating variables) and one slow state variable $\Vh$, corresponding to the feedback reference signal~\cite{izhikevich2007}. The speed at which $\Vh$ varies is $\eps \Delta_V$ with $\Delta_V(t)=0.183$\,mV\!$/$ms, where we extract the dimensionless small factor $\eps=10^{-2}$.

In contrast, the CC protocol holds $\Iext$, measuring the generated voltage $\Vcc$, thus, corresponding to an open-loop system, permitting e.g.\ the spiking seen in Fig~\ref{fig:fig1}B:
\begin{align}
\label{eq:CCSR}
\Iext &\approx\Ih,&\Vcc&\approx V,&\dotIh &= \eps \Delta_{I}.
\end{align}
The applied hold current $\Ih$ is varied slowly at speed $\eps \Delta_I$ with $\eps=10^{-2}$ and $\Delta_I=0.66\ldots0.75$\,pA$/$ms. General model \eqref{eq:gencond} with CC protocol \eqref{eq:CCSR} is also a slow-fast system with $1+N$ fast variables $(V,x_j)$ and $1$ slow variable $\Ih$. 
The gain $\gc$ in VC protocol \eqref{eq:VCSR} is limited by the imperfect conductance across the non-zero spatial extent of the membrane. Even though the conductance-based model \eqref{eq:gencond} is for the potential $V$ across the entire membrane and only $\Vh$ at the clamp is measured, we approximate $\Vh,\Vcc\approx V$ for the membrane potential, and $\Ih,\Ivc\approx\Iext$ for the external current in \eqref{eq:gencond}.

Following a classical multiple-timescale approach, we consider the $\eps=0$ limit of general model \eqref{eq:gencond} with VC protocol \eqref{eq:VCSR}, which corresponds to its $(1+\sum n_j)$-dimensional fast subsystem \eqref{eq:gencond}, with $\Iext=\gc(\Vh-V)$, where $\Vh$ is now treated as a parameter. For fixed $\Vh$ and voltages in the range $-80\ldots30$\,mV of interest, model \eqref{eq:gencond} with VC protocol \eqref{eq:VCSR} has only stable steady states (no limit cycles, that is, no neuronal spikes); see S1 Text for details.
For a fixed hold voltage $\Vh$, the steady states of system~\eqref{eq:gencond},\,\eqref{eq:VCSR}  with $\Iext=\gc(\Vh-V)$ satisfy the algebraic equations for $(V,x_j)_{j=1,\ldots,N}$
\begin{align}\label{eq:VCSRfast1eq}
\textstyle{\sum_j} I_j(x_j,V)&=\gc(\Vh-V),& x_j&=x_{\infty,j}(V),
\end{align}
where $\Ieq(V)=\sum_jI_j(x_{\infty,j},V)$ is the equilibrium current for fixed membrane potential $V$.
The solutions of algebraic system \eqref{eq:VCSRfast1eq} form a (1D) steady-state curve for model \eqref{eq:gencond} with VC protocol \eqref{eq:VCSR}, which is normally hyperbolic (transversally attracting) for $\eps=0$. 
For $\eps\neq0$ the increase of $\Vh$ with speed $\eps \Delta_V$ introduces a slow variation of all states $(V,x_j)$. Hence, $V$ and the  feedback current $\Ivc=\gc(\Vh-V)$ (as measured), are not at their steady-state values given by \eqref{eq:VCSRfast1eq}, but they are changing dynamically. This results in a difference between the measured curve $(\Ivc,\Vh)$ in Fig~\ref{fig:fig1}B and the $(I,V)$-values of the desired steady-state  $I-V$ curve given by \eqref{eq:VCSRfast1eq}. 

\begin{figure*}[!t]
\includegraphics[width=\textwidth]{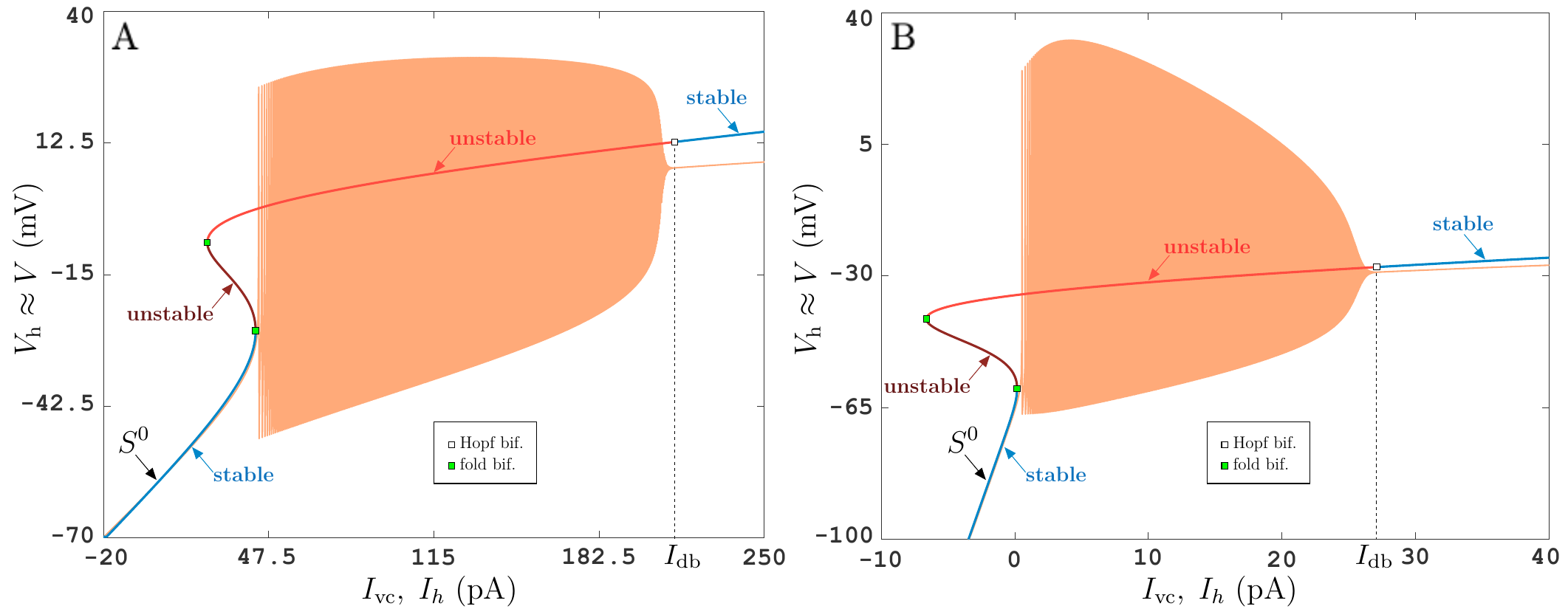}
\caption{{\bf VC and CC \textit{in silico}.} Protocol as described for Fig~\ref{fig:fig1}B applied to A: a \rev{class}-I Morris-Lecar neuron model~\eqref{eq:ML} \cite{morris1981} as example of excitatory cell, and B: a \rev{class}-I Wang-Buzs{\'a}ki neuron model~\eqref{eq:WB} \cite{wang1996} as example of inhibitory interneuron; see \eqref{eq:ML} and \eqref{eq:WB} for differential equations and Tables~\ref{tab:variables} and \ref{tab:variablesWB} for parameter values. The two-dimensional fast subsystem has a S-shaped steady-state curve satisfying steady-state conditions \eqref{eq:VCSRfast1eq}. The steady-state $I$--$V$ curve~\eqref{eq:VCSRfast1eq} and the (multi-color) S-shaped curve from the VC protocol \eqref{eq:VCSR} are indistinguishable throughout the range of input currents $\Ivc$. The orange curve resulting from the CC protocol is very close to $S^0$ and the VC protocol near its dynamically stable parts. \rev{See Figs\,F and G in S1 Text for numerical bifurcation diagrams.}
\label{fig:fig3}}
\end{figure*}
\begin{figure}[!h]
\centering
\includegraphics[width=0.5\columnwidth]{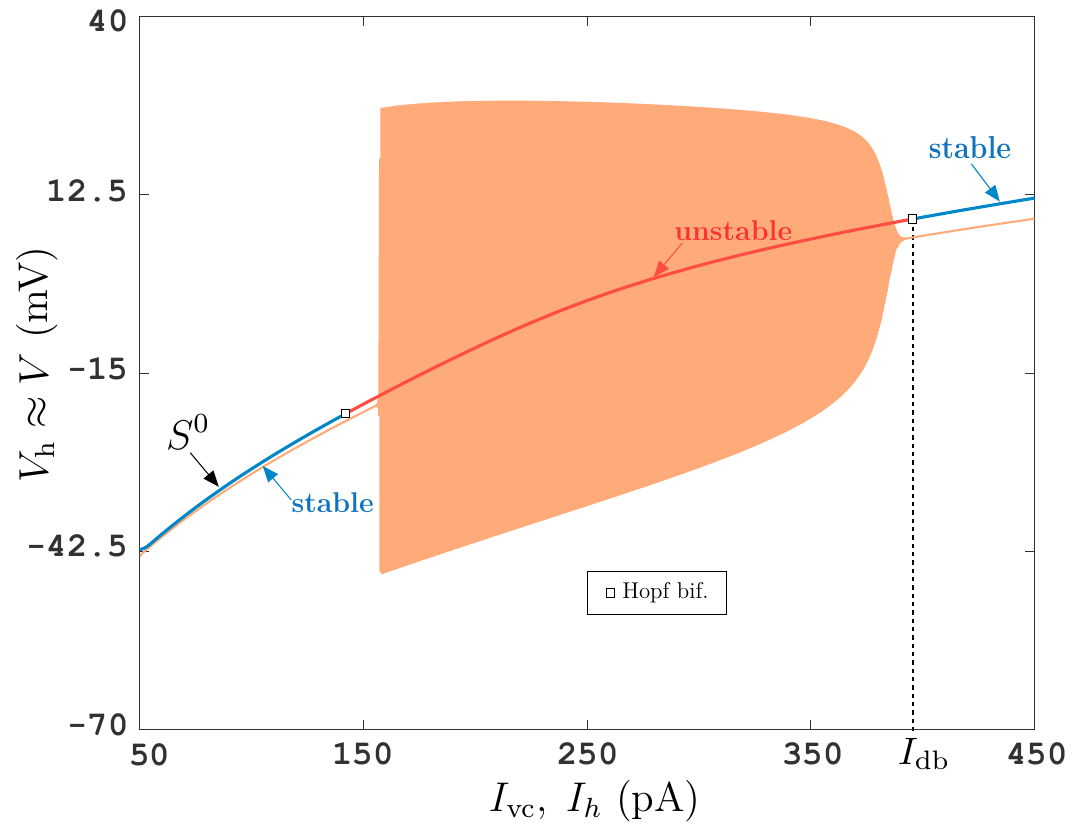}
\caption{VC and CC protocols  as described for Fig~\ref{fig:fig1}B applied to the Morris-Lecar model in a parameter regime where it behaves as a \rev{class}-II neuron model.  See \eqref{eq:ML} for differential equations and Table~\ref{tab:variables} for parameter values \rev{and Fig\,H in S1 Text for numerical bifurcation diagram}.}
\label{fig:fig4}
\end{figure}
Geometrical singular perturbation theory (GSPT) by Fenichel~\cite{fenichel1979} implies that after decay of initial transients every trajectory of the general model \eqref{eq:gencond} with VC protocol \eqref{eq:VCSR} satisfies the algebraic conditions \eqref{eq:VCSRfast1eq} for the steady-state curve up to order $\eps$. The first-order terms in $\eps$ are
\begin{equation}
\label{eq:close}
\begin{split}
\Ieq(V)-\Ivc(V)&\approx\Ieq'(V)\dot V\,\frac{\tau_{1/2}}{\ln2}\lesssim\Ieq'(\Vh)0.2\mbox{\,mV},
\end{split}
\end{equation}
where $\tau_{1/2}$ is the time for deviations from the transversally stable steady-state  $I-V$ curve~\eqref{eq:VCSRfast1eq} to decay to half of their initial value; see eq\,(7) and Fig\,C in S1 Text for details. We estimate $\tau_{1/2}$ from recovery transients after disturbances naturally occuring from imperfections in the voltage clamp during  VC runs as $\tau_{1/2}\lesssim0.075$\,s (see Fig\,CB in S1 Text), such that $\dot V\tau_{1/2}/\ln2\approx\dotVh\tau_{1/2}/\ln2\lesssim0.2$\,mV. Thus, the systematic bias between $\Ivcsm(\Vh)$ in Fig~\ref{fig:fig1}B and the true steady state curve $\Ieq(V)$  caused by dynamically changing $\Vh$ is below measurement disturbances.

Deviation estimate~\eqref{eq:close} can also be tested \emph{in silico}. Figs~\ref{fig:fig3} and~\ref{fig:fig4} emulate both VC and CC protocols with the Morris-Lecar model~\cite{morris1981} (see \nameref{sec:methods}, equation~\eqref{eq:ML}), a biophysical model typically used for excitatory neurons of either \rev{class}-I (Fig~\ref{fig:fig3}A) or \rev{class}-II (Fig~\ref{fig:fig4}) excitability, and with the Wang-Buzs{\'a}ki model (Fig~\ref{fig:fig3}B), which is a classical model of inhibitory interneuron \cite{wang1996} (see \nameref{sec:methods}, equation~\eqref{eq:WB}). Both models are of general form \eqref{eq:gencond}  with $j=n_j=1$. For the chosen parameter set (see S1 Text), the curves $(\Ivc(t),\Vh(t))$ and $(\Ieq(\Vh(t)),\Vh(t))$ are order $\eps$ ($\approx1$\%) apart. 

Consequently, time profile $(\Ivc(t),\Vh(t))$ follows closely the steady-state $I-V$  curve \eqref{eq:VCSRfast1eq} of \emph{stable} steady states of the fast subsystem \eqref{eq:gencond} with VC protocol $\Iext=\gc(\Vh-V)$, treating $\Vh$ as a parameter. \rev{This implies that faster ramp speeds are permissible when optimising trade-off between drift of cell properties and bias due to nonzero ramp speed. At high voltages the factor $\Ieq'(V)$ becomes large, such that estimate \eqref{eq:close} predicts larger deviations for large $\Vh$, as confirmed in Figs~\ref{fig:fig1}B, \ref{fig:fig2}, AB, AC and AD in S1 Text.}

We now connect the steady-state  $I-V$ curve \eqref{eq:VCSRfast1eq} of the VC protocol to a curve of fast-subsystem equilibria of the CC protocol, which is in part unstable. To this end we recast the VC protocol in the form of a CC protocol with non-constant current ramp speed $\eps\Delta_{I,\mathrm{vc}}(t)$ and disturbances: the smoothed time profile $\Ivcsm(\Vh(t))$ in Fig~\ref{fig:fig1}B of the VC run (thick, in blue/brown/red) equals the raw-data measured time profile $\Ivc(t)$  (thin blue curve with fluctuations) plus disturbances $\etavc(t)$, defined by $\etavc(t)=\Ivcsm(\Vh(t))-\Ivc(t)$. After smoothing, the derivative $\Ivcsm'(\Vh)$ w.r.t.\ $\Vh$ is moderate ($\lesssim20$\,pA$/$mV in modulus at its maximum near $I_\mathrm{db}$), such that during the VC protocol the external current $\Iext$ satisfies
\begin{equation}
\label{eq:VCSRI}
    \Iext \approx\Ivcsm+\etavc,\ 
    \dotIvcsm = \eps \Delta_{I,\mathrm{vc}}(t),
\end{equation}
where $\Delta_{I,\mathrm{vc}}(t)=\Ivcsm'(\Vh(t))\Delta_{V}$. Hence, $\Delta_{I,\mathrm{vc}}(t)\approx\Ieq'(\Vh(t))\Delta_V$, with upper bound $\max_t|\Delta_{I,\mathrm{vc}}(t)|\lesssim3.7$\,pA$/$ms in the range of Fig~\ref{fig:fig1}B. Thus, $\Ivcsm$ is indeed still slow.
So, except for disturbances $\etavc(t)$, the external current $\Iext$ is slowly varying according to a CC protocol with slowly time-varying speed $\eps \Delta_{I,\mathrm{vc}}(t)$, such that the VC protocol \eqref{eq:VCSR} is equivalent to the CC protocol \eqref{eq:VCSRI}  with disturbances $\etavc$.

This means that general model~\eqref{eq:gencond} with current $\Iext$ given in \eqref{eq:VCSRI} with zero disturbances ($\etavc=0$) is a model for a CC protocol with driving current $\Ivcsm$, in contrast to the model with open-loop CC protocol \eqref{eq:gencond},\,\eqref{eq:CCSR}. Both models have the same fast subsystem \eqref{eq:gencond} when setting $\eps=0$ (i.e., for constant input current) and identifying $\Ih$ and $\Ivcsm$. The respective fast-subsystem steady states $(V,x_j)_{j=1,\ldots,N}$ satisfy
\begin{equation} \label{eq:VCSRIfasteq}
\begin{split}
    \textstyle{\sum_j} I_j(x_j,V)&=\Ih(=\Ivcsm),\\ x_j&=x_{\infty,j}(V).
\end{split}
\end{equation}
However, the two models differ by the nature of their respective slow variables $\Ih$ and $\Ivcsm$: $\Ih$ is an externally applied hold current for the open-loop CC protocol \eqref{eq:CCSR}, while $\Ivcsm$ is a measured (and smoothed) current from the closed-loop feedback control $\gc(\Vh-V)$ of the voltage source for \eqref{eq:VCSRI}. Thus, while the S-shaped steady-state curve $(\Ieq(V),V)$ is identical for both models, it contains large unstable segments as a steady-state curve of open-loop CC protocol \eqref{eq:gencond},\,\eqref{eq:CCSR}, while it is always stable as a steady-state curve of closed-loop VC protocol \eqref{eq:gencond},\,\eqref{eq:VCSRI}. The change in stability is caused by the disturbances $\etavc$, which are current adjustments generated by the feedback term in VC protocol \eqref{eq:VCSR}, $\Ivc=\gc(\Vh-V)$. Along most of the curve $(\Ivcsm(\Vh),\Vh)$ the $\etavc$ are small fluctuations such that $\Ivcsm(\Vh)\approx\Ivc(\Vh)$ and the feedback is approximately non-invasive \cite{sieber2008a}. Estimate~\eqref{eq:close} ensures that the measurements $\Ivc(V)$ stay close to $\Ieq(V)$. Therefore, we can conclude that the VC protocol \eqref{eq:VCSR} with slowly varying feedback reference signal $\Vh$ reveals the entire family of steady states of a neuron (\rev{class} I or II) with constant external current $\Iext$, both stable (observable) and unstable (non-observable, hidden). Consequently, the N-shaped $I$-$V$ relations for \rev{class}-I neurons reported in \cite{schwindt1981,fishman1969,vervaeke2006} equal S-shaped steady-state bifurcation diagrams for these neurons with respect to $\Iext$. They are tractable with a VC protocol where the current $\Iext$ is a sufficiently slowly varying feedback current with sufficiently small fluctuations $\etavc$. In particular, this allows us to detect and pass through fold bifurcations directly in the experiment. 

\begin{figure*}[!t]
\centering
\includegraphics[width=\textwidth]{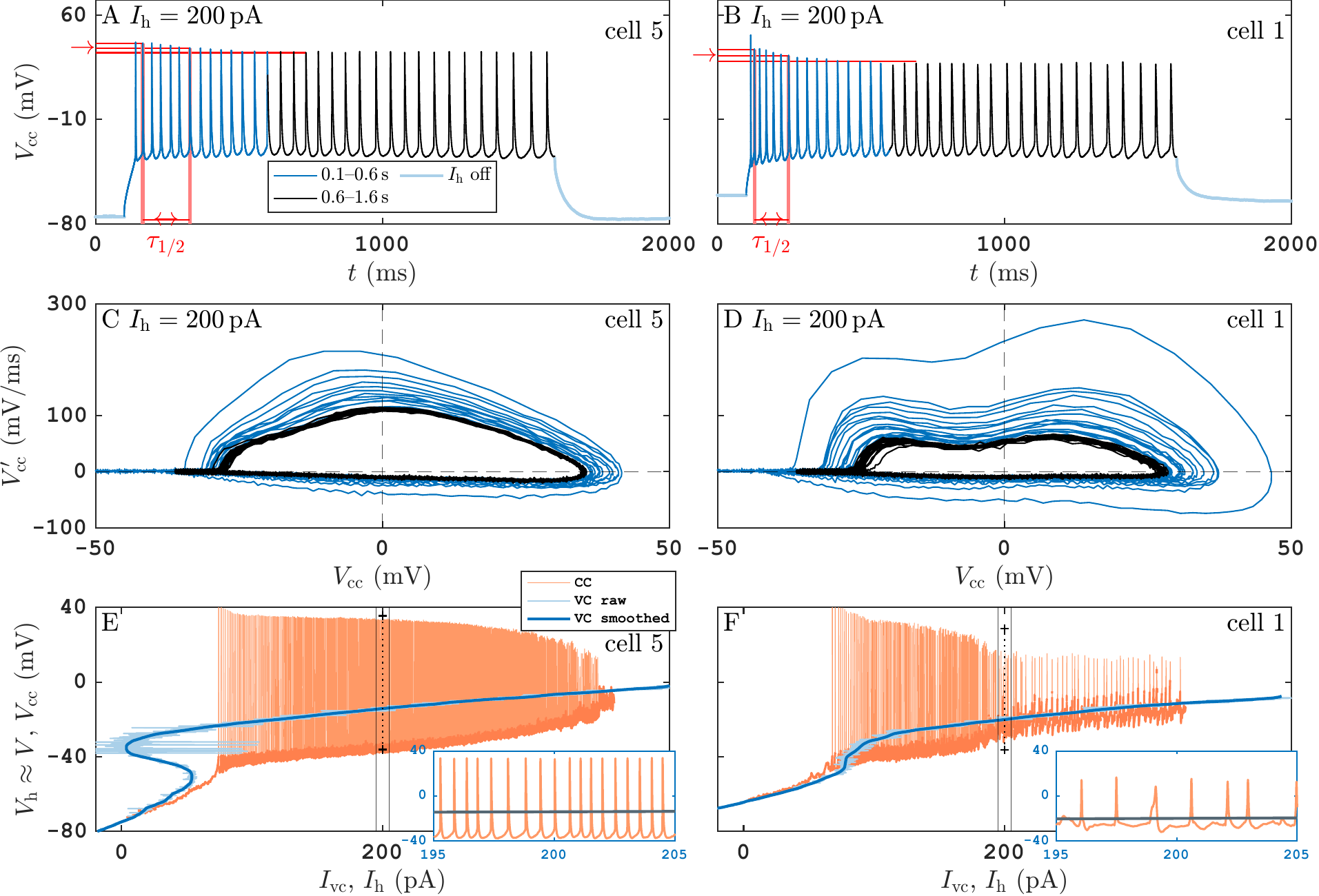}
\caption{Periodic spiking responses to step-current protocols for PY cells 5 (left column) and 1 (right column).  Panels A, B: time profiles of voltage responses $\Vcc(t)$ from a current-clamp stimulation with a step to constant hold current $\Ih=200$ pA from $0.1$ to $1.6$\,s ($0.1$--$0.6$\,s in blue, $0.6$--$1.6$\,s in black). Red markings show how half-decay time $\tau_{1/2}$ is extracted from voltage maxima during transients. Panels C, D: $(\Vcc,\Vcc')$ phase-plane projection the time series from panels A, B, using a one-step finite-difference approximation of $\Vcc^{\prime}(t)$ with color code matching panels A, B to distinguish transients and steady-state spiking. Panels E, F: reproduction of Figs~\ref{fig:fig1}A and AA in S1 Text without bifurcation or stability markings, respectively. The value $\Ih=200$\,pA (vertical dotted black line) and a $10$\,ms window around it (vertical solid black lines) are highlighted. The insets show a zoom into this $10$\,ms window around $\Ih=200$\,pA of panels E, F. Black crosses in panels E, F are minimum and maximum of steady-state spiking from panels A, B for comparison.}
\label{fig:fig5}
\end{figure*}   
In contrast, during the CC open-loop protocol \eqref{eq:CCSR}, when applying a slowly varying electrical current, the neuron dynamically transitions between its observable rest states and its observable (dynamically stable) spiking states (see Fig~\ref{fig:fig1}B orange time profile). The speed of variation $\eps \Delta_I\lesssim 0.75\times10^{-3}$\,pA$/$ms is such that $\Iext$ varies by $1$\,pA or less per spiking period. For a transient decay analysis we  stimulate the neuron with larger current steps during  a calibration phase before executing CC protocol runs. Fig~\ref{fig:fig5}A and \ref{fig:fig5}B show the response to such a current step to $\Iext=200$\,pA for cells $5$ and $1$.  We observe that the transients in the step current response such as in Fig~\ref{fig:fig5}A and \ref{fig:fig5}B in the stable spiking region have a half-time for decay toward the stable spikes
\begin{align}
  \label{eq:decayhalftime}
  \tau_{1/2}\lesssim0.2\mbox{\,s.}
\end{align}
This half-time $\tau_{1/2}$ enters estimates for the bias caused by varying the hold current $\Ih$ dynamically with speed $\eps\Delta_I$. For example, the estimate for the bias in the minimum voltage $V_{\min}$ of the spike equals to first order in $\eps$ 
\begin{align}\label{eq:biasvmin}
\bias(V_{\min})\approx V_{\min}'(\Ih)\times \frac{\mathrm{d}}{\mathrm{d}t}\Ih(t)\times\frac{\tau_{1/2}}{\ln2}.    
\end{align}
We can see in Fig~\ref{fig:fig5}E and \ref{fig:fig5}F that $V_{\min}(\Ih)$ changes in the spiking region with about $8$\,mV per $120$\,pA, so $V_\mathrm{min}'(I)\approx 0.07$\,mV$\!/$pA. The current changes with $\mathrm{d}I/\mathrm{d}t\approx7.5$\,pA$/$s. Thus, to first order in $\eps$, the error then amounts to:
$$\bias(V_{\min})\approx V_{\min}'(\Ih)\times \frac{\mathrm{d}}{\mathrm{d}t}\Ih(t)\times\frac{\tau_{1/2}}{\ln2}\approx0.14\mbox{\,mV.}$$ Hence, the effect from changing the current $\Ih$ dynamically is small (below visibility in bifurcation diagrams such as Fig~\ref{fig:fig1}B or Fig~\ref{fig:fig5}E and \ref{fig:fig5}F. The linear bias estimate grows to infinity when the amplitude $V_{\max}(\Ih)$ envelope assumes a square-root like shape and $\tau_{1/2}$ goes to infinity, as is the case at Hopf bifurcations.

Thus, combining the VC protocol \eqref{eq:VCSRI} for varying $\Ivcsm$, and the CC protocol \eqref{eq:CCSR} for varying $\Ih$, enables us to interpret the data sets from both protocols in Fig~\ref{fig:fig1}B as a bifurcation diagram including unstable states. 
\paragraph{Drift and intermittent dropouts}
In Fig~\ref{fig:fig5}E and \ref{fig:fig5}F the black markers ($+$) at $\Ih=200$\,pA  mark the maximum and minimum values of the voltage response (after transient) to the step-current stimulation, as shown in panels A and B. They indicate that for PY cell 5 there is quantitative agreement between the response to the CC protocol with slowly-varying applied current $\Ih(t)$ and the response to the current step. There is less agreement for PY cell 1, between step response and the response to the CC protocol with slow variation of $\Ih$. The time profile suggests that there are either different possible spiking responses and, hence, more bifurcations in the dynamic protocol, such as period-doubling bifurcations, or the condition of the PY cell has drifted between step response calibration and CC ramp.

Furthermore, the distances between $(\Ivcsm(t),\Vh(t))$ in the VC run and $(\Ih(t),\Vcc(t))$ in the CC run for the same cell are visibly larger than predicted by eq~\eqref{eq:close} along parts of the curve corresponding to dynamically stable stationary states. This is due to the natural \textit{drift} of the neuron's physiological properties as it changes conductance properties (e.g., temperature, degradation,\ldots). \rev{In S1 Text we show that} over time.

Finally, for the experimental curves presented in Figs~\ref{fig:fig1}B, \ref{fig:fig1}C, \ref{fig:fig2} and A in S1 Text, the disturbances $\etavc$ are not small in some unstable parts of the reported steady-state curve (e.g., near $\Vh=-30$\,mV in Fig\,\ref{fig:fig1}B), caused by imperfect voltage clamping across the membrane. Fig\,CB in S1 Text shows a zoom demonstrating that the observed current spikes are indeed intermittent dropout events. Detailed simulations using Morris-Lecar model~\eqref{eq:ML}, shown in S1 Text Section 3, reproduce these dropouts closely, if adds short current spikes and small-amplitude white noise to \eqref{eq:ML}.

\section{Discussion}
Tracking non-observable (hidden) states and their stability boundaries in experimental settings bridges the gap between real-world phenomena and nonlinear science. Specifically, closed-loop control methods with slow variations of feedback reference signals enable to dissect the underlying states of multi-scale complex systems. Following this strategy, we demonstrate the feasibility of tracking a family of neuronal steady states (stable and unstable) via variations of reference signal and reparameterizing the obtained curve by feedback current. The method is versatile in that it applies to different cell types, both pyramidal (PY) neurons (like in Fig~\ref{fig:fig1}), of excitability \rev{class} I or II (Figs~\ref{fig:fig3}A and~\ref{fig:fig4}), and inhibitory interneurons (Fig~\ref{fig:fig3}B).

Our analysis and the resulting estimate for the nonzero-speed induced bias during the VC protocol in \eqref{eq:close} indicates that for the steady states detected in VC ramps our ramp speed (less than $3$\,mV$\!/$s) does not create a bias that is notable when compared to drift of cell properties over time or fluctuations caused by external effects (such as the dropouts visible in Fig~\ref{fig:fig1}B). For the CC protocol and parameter areas were periodic spiking is stable, transient analysis for step current responses results in estimates such as \eqref{eq:biasvmin}, where bias caused by nonzero current ramp speed ($\approx7.5$\,pA$/$s) is also negligible. This implies that faster ramp speeds are permissible when optimising trade-off between drift of cell properties and bias due to nonzero ramp speed.
\rev{The single-ramp nature of the CC protocol leaves the  question of stability open for some parts of the steady-state branch. These results can be improved by varying the hold current $\Ih$ non-monotonically and be switching from VC to CC protocol at selected parts of the steady-state branch determined by the VC protocol.}

These results can be also extended to more complex neuronal states (e.g. limit cycles) by combining two recent advances, namely: 1. \textit{dynamic-clamp} electrophysiology\cite{sharp1993,marder1996,chizhov2014,kirst2015,ori2020,pfeiffer2022}, which allows for a two-way real-time communication between a neuronal tissue (e.g. neuron, network or neuronal sub-processes) and a computer simulation of a component of the neuronal tissue. This introduces the full range of feedback control theory and engineering techniques into electrophysiology; 2. \textit{Control-based continuation for experiments (CBCE}, which combines feedback control theory and pseudo-arclength numerical continuation for tracking solution branches of nonlinear systems directly from noisy experimental data~\cite{sieber2008a}. Indeed, CBCE method has been successfully applied in various experimental systems, for instance, in mechanical vibration and buckling experiments \cite{RENSON2019449}, pedestrian flow experiments \cite{panagiotopoulos2022control}, cylindrical pipe flow simulations \cite{willis_duguet_omelchenko_wolfrum_2017} and feasibility studies for synthetic gene networks \cite{de2022control,blyth2023numerical}. Noteworthy, the methodology presented in this work can be seen as a special case of CBCE. Future work will focus on employing the full CBCE to track more complex neuronal states\rev{, for instance, unstable spiking states in single neurons or unstable traveling-wave states in neural tissue using micro-electrode arrays (MEA)}. This is crucial as it will allow us to validate computational models by comparing their numerical bifurcation diagrams with experimental ones\rev{; it will also help obtaining better model fitting based on both voltage-clamp and current-clamp data}. This approach will have direct impact in experimental labs enabling biologists to have access to novel states with which to carry out new experimental paradigms. We envisage this extended protocol could be used to develop next-generation closed-loop deep-brain stimulation devices to treat certain brain diseases such as epilepsy~\cite{breakspear2006}, where it could help monitor control the excitability threshold of key neural populations in real time.

\section{Materials and Methods}
\label{sec:methods}
\subsection{Ethics Statement}
All animal treatments were authorized by the Sechenov Institute of Evolutionary Physiology and Biochemistry Bioethics Committee (protocol no. 1-7/2022, 27 January 2022) and adhered to the European Community Council Directive 86/609/EEC.

\subsection{Protocols}
Male Wistar rats were used in this study (age P21, N=8 animals). For the bifurcation diagrams (see Figs~\ref{fig:fig1}B and A in S1 Text), 4 of these 8 rats were used. PY Cells 1-3 were recorded from different animals, PY cells 4-5 were recorded from the same animal. To study the effects of hysteresis (see Fig~\ref{fig:fig2}) 20 neurons were recorded from the 4 remaining animals (221222, 230110, 230111, 230112). The control group were 6 neurons from 2 rats (230110, 230112). The QX group were 8 neurons from 4 rats (221222, 230110, 230111, 230112). The QX+Cd group were 6 neurons from 4 rats (221222, 230110, 230111, 230112).

The use and handling of animals was performed in accordance with the European Community Council Directive 86/609/EEC. Horizontal 300-{\textmu}m-thick brain slices were prepared as described in our previous studies~\cite{amakhin2021}. The slices contained the hippocampus and the adjacent cortical regions and were kept in the artificial cerebrospinal fluid (ACSF) with the following composition (in mM): 126 NaCl, 24 NaHCO$_3$, 2.5 KCl, 2 CaCl$_2$, 1.25 NaH$_2$PO$_4$, 1 MgSO$_4$, 10 glucose (bubbled with 95\% O2/5\% CO$_2$ gas mixture). All the listed chemicals were purchased from Sigma-Aldrich (St. Louis, MO, USA).

We performed the whole-cell patch-clamp recordings of the principal neurons in the entorhinal cortex. Neurons within the slices were visualized using a Zeiss Axioscop 2 microscope (Zeiss, Oberkochen, Germany), equipped with a digital camera (Grasshopper 3 GS3-U3-23S6M-C; FLIR Integrated Imaging Solutions Inc., Wilsonville, OR, USA) and differential interference contrast optics.

Patch pipettes were produced from borosilicate glass capillaries (Sutter Instrument, Novato, CA, USA) and filled with one of the following pipette solutions. A potassium gluconate-based pipette solution had the following composition (in mM): 136-K-Gluconate, 10 NaCl, 10 HEPES, 5 EGTA, 4 ATP-Mg, 0.3 GTP. For control experiments described in Section\,SI4, the intracellular block of the voltage-gated sodium channels was required, for which we used the solution with added QX314 (Alamone labs, Jerusalem, Israel), which had the following composition (in mM):  130-K-Gluconate, 10 HEPES, 6 QX314, 6 KCl, 5 EGTA, 4 ATP-Mg, 2 NaCl, 0.3 GTP. The pH of both solutions was adjusted to 7.25 with KOH. The resistance of filled patch-pipettes was 3-4 M$\Omega$.

A Multiclamp 700B (Molecular Devices, Sunnyvale, CA, USA) patch-clamp amplifier, a NI USB-6343 A/D converter (National Instruments, Austin, TX, USA) and WinWCP 5 software (SIPBS, UK) were used to obtain the electrophysiological data. The recordings were lowpass filtered at 10 kHz and sampled at 20-30 kHz. The access resistance was less than 15 M$\Omega$ and remained stable during the recordings. The liquid junction potential was not compensated for. The flow rate of ACSF in the recording chamber was 5 ml/min. The recordings were performed at 30$^\circ$C.

Specifically for the voltage-clamp protocol. We note the CV-7B headstage has four different feedback resistors (Rf): 50 M$\Omega$, 500 M$\Omega$, 5 G$\Omega$, and 50 G$\Omega$. The Rf determines the maximum currents that can be recorded or injected. In voltage-clamp mode it is generally recommended to use the largest possible value of Rf (larger Rf results in less noise), though high values can result in current saturation. Since in our preparation the electrical currents varied between 50-2000 pA (several nA for the potassium ion currents at positive holding potentials), we chose Rf= 500 M$\Omega$  (i.e. feedback gain $g_c = 2$\,nS).

We applied first VC and then CC protocol to $5$ neurons in the entorhinal cortex of $4$ male Wistar rats \cite{amakhin2021}.  We first performed the VC neuronal recordings varying the hold voltage $\Vh$ from $-80$\,mV to $+30$\,mV slowly with $\dotVh=1.83$\,\mbox{mV\!$/$s}, while measuring current, called $\Ivc$ in Figs~\ref{fig:fig1}B and \ref{fig:fig5}C. Subsequently, for the CC recordings we first determine the minimal injected current required to induce a depolarization block of action potential generation ($I_{\mathrm{db}}$ in Fig~\ref{fig:fig1}B, upper limit of current input where firing occurs).
Then we injected hold current $\Ih$, gradually increasing from $0$\,pA to $I_{\mathrm{db}}$ during $60$\,s, such that $\dotIh$ is in the range $6.6\ldots7.4$\,pA$/$s for the $5$ neurons, while recording voltage $\Vcc(t)$. See S1 Text for further checks (e.g., for hysteresis).

\subsection{Computational models}
\label{sec:models}
We applied the VC and the CC protocols, with slow variations in the feedback reference signal and in the applied current, respectively, to the classical neuron model adapted to many contexts, namely the Morris-Lecar model~\cite{morris1981} (ML), and to a standard neuron model that was specifically designed to account for inhibitory neural activity, namely the Wang-Buzs{\'a}ki model~\cite{wang1996} (WB). Parameter values are given in Tables~\ref{tab:variables} and~\ref{tab:variablesWB} below, respectively.

The ML equations are as follows
\begin{equation}\label{eq:ML}
\begin{split}
C\dot{V} &= - g_\mathrm{L}(V-V_\mathrm{L})-g_{Ca}m_{\infty}(V)(V-V_{Ca})-g_{K}w(V-V_{K})+g_c(V_{\mathrm{h}}-V),\\
\dot{w} &= \phi\frac{w_{\infty}(V)-w}{\tau_{w}(V)},
\end{split}
\end{equation}
with the following voltage-dependent (in)activation and time-constant functions: 
\begin{equation}\label{eq:inact}
\begin{split}
m_{\infty}(V) &= 0.5(1+\tanh((V-V_1)/V_2)),\\
w_{\infty}(V) &= 0.5(1+\tanh((V-V_3)/V_4)),\\
\tau_{w}(V)   &= \cosh((V-V_3)/(2V_4))^{-1}.
\end{split}
\end{equation}
To obtain Figs~\ref{fig:fig3}A and~\ref{fig:fig4}, we have used the following parameter values.
\begin{table}[!h]
\begin{center}
	\caption{Parameter values for the Morris-Lecar model~\eqref{eq:ML}.}
	\label{tab:variables}
    \setlength{\tabcolsep}{1.2ex}
	\begin{tabular}{lccccccccccccc}
		\hline\\[-0.25cm]
		parameter&$C$ & $g_L$ & $V_L$ & $g_{Ca}$ & $V_{Ca}$ & $g_K$ & $V_K$ & $g_c$ &
            $\phi$ & $V_1$& $V_2$ & $V_3$ & $V_4$ \\[0.25ex]
        unit&pF & nS & mV & nS & mV & nS & mV & nS &
         & mV & mV & mV & mV\\[0.25ex]
        value \rev{class}-I&  20 & 2 & -60 & 4.0 & 120 & 12 & -84 & 40 & 0.067 & -1.2 & 18 & 12 & 17.4\\[0.25ex]
        value \rev{class}-II&  20 & 2 & -60 & 4.4 & 120 & 12 & -84 & 150 & 0.040 & -1.2 & 18 & 2 & 30.0
	\end{tabular}
\end{center}
\end{table}
Finally, for both the VC protocol with slow variation, and the CC protocol with slow variation, the speed of the variation was chosen to be equal to $\eps=0.01$. Note that $g_c$ can be decreased to 7, which is in the same order of magnitude as the experimental one. Moreover, the capacitance chosen for the model simulations is on the same order of magnitude as observed in the experiments. Nevertheless, the key point to note is that we are not aiming for quantitative agreement since this is a conductance-based phenomenological model.

The WB equations are as follows
\begin{equation}\label{eq:WB}
\begin{split}
C\dot{V} &= - g_\mathrm{L}(V-V_\mathrm{L})-g_{Na}m_{\infty}^3(V)h(V-V_{Na})-g_{K}n^4(V-V_{K})+g_c(V_{\mathrm{h}}-V),\\
\dot{h} &= \phi\frac{h_{\infty}(V)-h}{\tau_{h}(V)},\\
\dot{n} &= \phi\frac{n_{\infty}(V)-n}{\tau_{n}(V)},
\end{split}
\end{equation}
with the following voltage-dependent (in)activation and time-constant functions: 
\begin{equation}\label{eq:inactWB}
\begin{split}
\alpha_m(V) &= 0.1(V+35)/(1-\exp(-0.1(V+35)))\\
\beta_m(V)  &= 4.0\exp(-0.0556(V+60))\\
\alpha_h(V) &= 0.07\exp(-0.05(p+58))\\
\beta_h(V)  &= 1/(1+\exp(-0.1(V+28)))\\
\alpha_n(V) &= 0.01(V+34)/(1-\exp(-0.1(V+34)))\\
\beta_n(V)  &= 0.125\exp(-0.0125(V+44))\\
p_{\infty}(V) &= \alpha_p(V)/(\alpha_p(V)+\beta_p(V)),\;p\in\{m,h,n\},\\
\tau_p(V) &=
1/(\alpha_p(V)+\beta_p(V)),\;p\in\{m,h,n\}\end{split}
\end{equation}
To obtain Fig~\ref{fig:fig3}B, we have used the following parameter values.
\begin{table}[!h]
\begin{center}
	\caption{Parameter values for the Wang-Buzs{\'a}ki model~\eqref{eq:WB}.}
	\label{tab:variablesWB}
    \setlength{\tabcolsep}{1.2ex}
	\begin{tabular}{lccccccccc}
		\hline\\[-0.25cm]
		parameter&$C$ & $g_L$ & $V_L$ & $g_{Na}$ & $V_{Na}$ & $g_K$ & $V_K$ & $g_c$ &
            $\phi$ \\[0.25ex]
        unit&pF & nS & mV & nS & mV & nS & mV & nS &\\[0.25ex]
        value&  1 & 0.1 & -65 & 35.0 & 55 & 9 & -90 & 20 & 5 
	\end{tabular}
\end{center}
\end{table}
Finally, for both the VC protocol with slow variation, and the CC protocol with slow variation, the speed of the variation was chosen to be equal to $\eps=0.01$.

\section*{Data and code availability}
All presented data, as well as, codes to process them, are available via the following Github repository:\newline\href{https://github.com/serafimrodrigues/Amakhin-etal-2025/}{https://github.com/serafimrodrigues/Amakhin-etal-2025/}.

\paragraph*{S1 Text.}
\label{S1_text}
Supplementary text and Supplementary Figures.

\section*{Acknowledgments}
SR acknowledges the grant PID2023-146683OB-100 funded by\newline MICIU/AEI /10.13039/501100011033 and by ERDF, EU. Additionally, SR acknowledges support from Ikerbasque Foundation and the Basque Government through the BERC 2022-2025 program and by the Ministry of Science and Innovation: BCAM Severo Ochoa accreditation CEX2021-001142-S / MICIU / AEI / 10.13039/501100011033. Moreover, SR acknowledges the financial support received from BCAM-IKUR, funded by the Basque Government by the IKUR Strategy and by the European Union NextGenerationEU/PRTR, as well as, support of SILICON BURMUIN no. KK-2023/00090 funded by the Basque Government through ELKARTEK Programme.

\bibliographystyle{plain}
\bibliography{refs}

\newpage

\renewcommand{\thesection}{S\arabic{section}}
\renewcommand{\thefigure}{\Alph{figure} in S1 Text}
\renewcommand{\theequation}{S\arabic{equation}}
\setcounter{equation}{0}  
\setcounter{section}{0}  
\setcounter{figure}{0}  

\begin{center}
	\fontsize{20}{24}\bfseries\scshape{S1 Text}
\end{center}

\section{Experimental bifurcation diagrams for Pyramidal (PY) cells 1--5, including current and voltage clamp ramp data}
\begin{figure*}[ht]
\centering
\includegraphics[width=0.8\columnwidth]{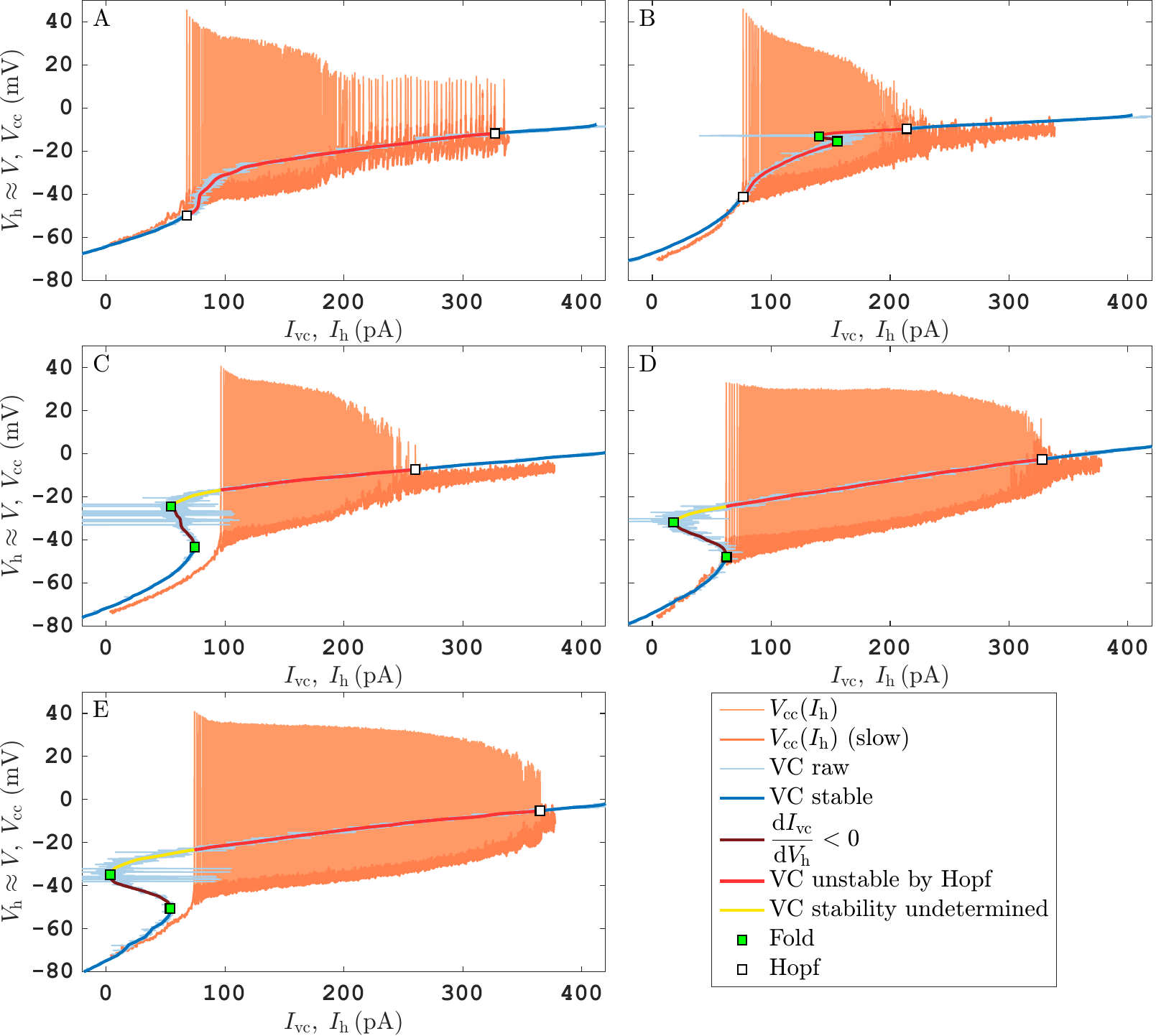}
\caption{Experimental bifurcation diagrams for PY cells 1--5 (panel E identical to Fig\,1B of main manuscript). Protocols and color codes identical to Fig\,1B in main manuscript.}
\label{fig:figS1}
\end{figure*}   
Fig~\ref{fig:figS1} presents the recordings for VC and CC protocol for all PY cells (1--5) in the same way as Fig\,1B in the main manuscript (which was only for PY cell~5). The graphs superimpose the observations of our VC protocol with slow variation of the feedback reference signal and the CC protocol with slow variation of the externally applied currents. The graphs confirm (as does Fig\,1C from the main manuscript) that \rev{(Hodgkin excitability) class}-II properties of the neurons vary between individual cells. We also observe that in all PY cells for $\Ih>I_\mathrm{db}$ and for PY cell 3 for small $\Ih$ the distance between the stable steady states obtained from the VC protocol and the corresponding trajectory segments obtained from the CC protocol is larger than the bias caused by dynamic variation of $\Vh$ and $\Ih$ respectively, based on estimate (5) of the main text. We do not have a complete physiological explanation for this mismatch, but we hypothesise that this is due to the natural physiological drift of underlying neuronal processes as one performs successive recordings on the same PY cell (see also Fig\,2 of the main manuscript for a comparison between successive ramp-up and ramp-down on the same PY cell).

\section{Spiking frequency and interspike intervals during CC protocol}
\begin{figure*}[!t]
\centering
\includegraphics[width=1\columnwidth]{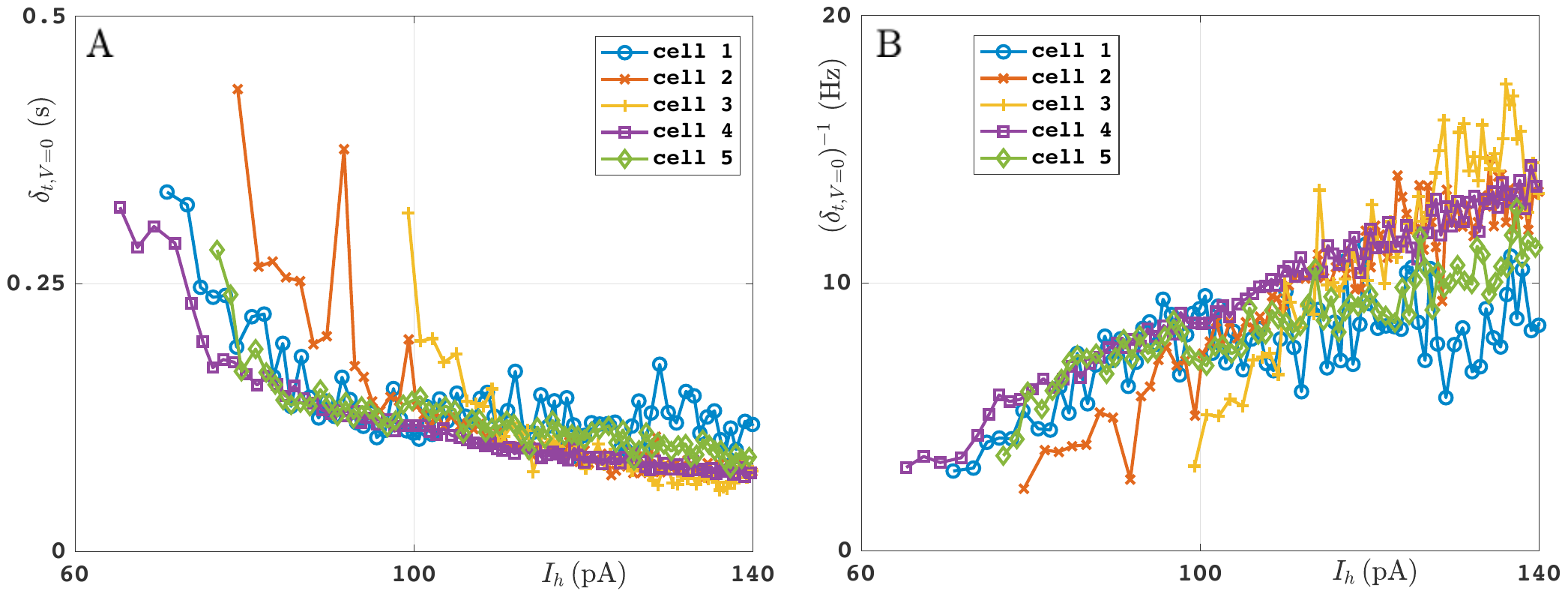}
\caption{Interspike intervals  in seconds (panel A) and frequency in Hz (panel B) for the spiking responses during the CC protocols with slowly-varying applied current near the low-current limit of the spiking region.}
\label{fig:figS2}
\end{figure*}

A characteristic for the class of excitability is the dependence of the spiking frequency on the external current $\Iext$ as $\Iext$ approaches the lower limit of the spiking region, with a drop in frequency expected for \rev{class}-I neurons. Fig~\ref{fig:figS2} presents the interspike interval as a function of the hold current $I_\mathrm{h}$ in panel A, and the frequency against $I_\mathrm{h}$ in panel B for the 5 PY cells for which experimental bifurcation diagrams are shown in Fig\,1 of the main manuscript and Fig~\ref{fig:figS1}. 
We detected crossing times for $\Vcc(t)$ through the threshold $V=0$ from below (or equivalently Poincar{\'e} map, at $V=0$). As visible in Fig~\ref{fig:figS2}, even though the 5 PY cells have different steady-state curves (obtained with the VC protocol), their frequency behaviours during the CC protocol are qualitatively similar with drops from $\approx15$\,Hz to $\lesssim5$\,Hz for all PY cells. Hence, the frequency drop does not allow us to conclusively distinguish between different excitability classes for the present PY cells. However, the steady-state bifurcation diagrams obtained through VC protocols show that PY cells 3-5 exhibit an experimental bifurcation diagram that is compatible with \rev{class}-1 excitability, while PY cells 1 and 2 have \rev{class}-2 compatible steady-state curves.

\section{Details of intermittent fluctuations around unstable steady states}\label{sec:fluctuations}
When performing the VC protocol with slow variation of the feedback reference signal we observe fluctuations around the states that would be unstable in the CC protocol. The highlighted ellipse in Fig\,CA in S1 Text shows a prominent example of large-scale fluctuations. We observe this phenomenon in all PY cells. Fig\,CB in S1 Text shows details of the typical time profile of these fluctuations in the measured feedback current $\Ivc = g_c (V_h- V)$ in blue. In the background Fig\,CB in S1 Text shows a neuronal firing pattern  (in orange) around the same currents to give a visual comparison of fluctuation amplitude and frequency with the natural neuronal firing patterns. We also compute the moving average of $\Ivc(t)$ (dashed blue), which shows the equilibrium level of $\Ivc$. We observe that between the large fluctuations the feedback current $\Ivc$ stabilises to this equilibrium level with a consistent rate of decay and consistently smaller levels of smaller fluctuations. We used the observed decay rate toward the mean to estimate the decay half-time $\tau_{1/2}$ in eq.\,(5) of the main manuscript.

\begin{figure*}[!t]
\centering
\includegraphics[width=1\columnwidth]{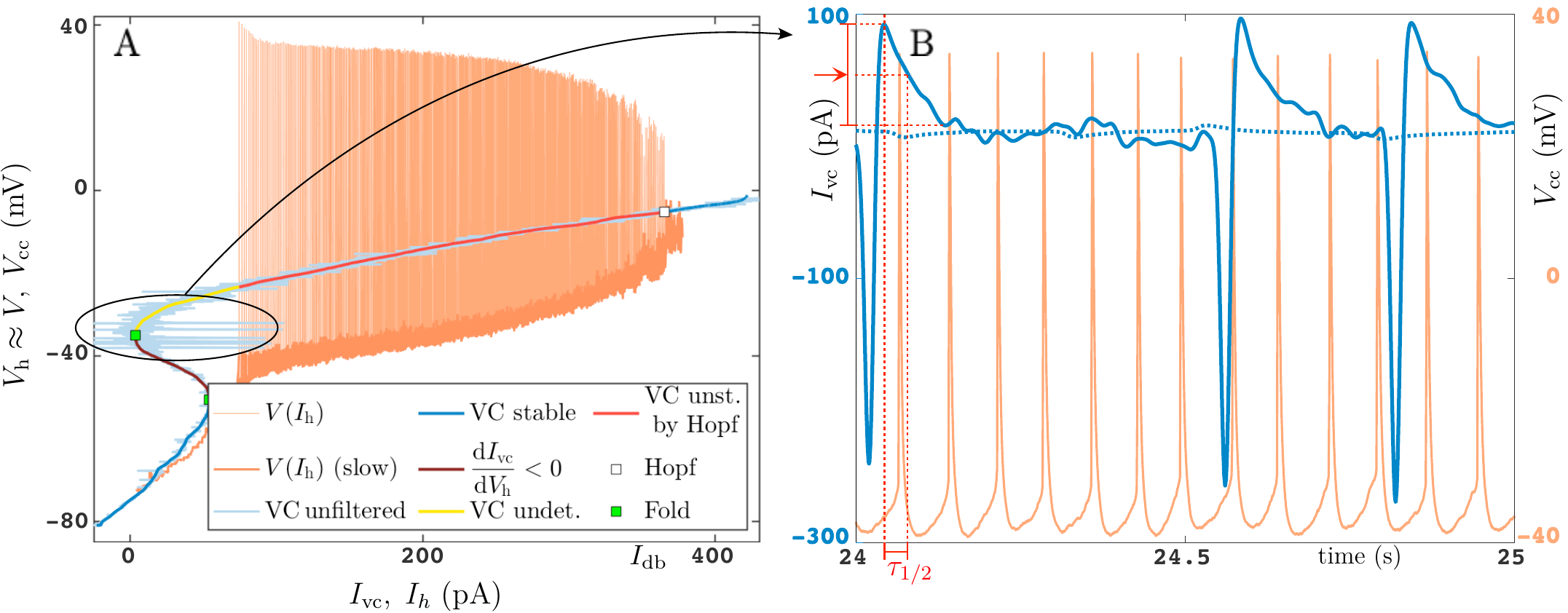}
\caption{Experimental intermittent fluctuations. A: Reproduction of Fig\,1B of main manuscript with the region of intermittent fluctuations highlighted by a black ellipse. B: Fluctuation of the current $\Ivc$ from the VC protocol, with apparent intermittent fluctuation (blue), compared with the voltage spikes obtained from the CC protocol (orange). Red dashed lines illustrate how we compute $\tau_{1/2}$.}
\label{fig:figS3}
\end{figure*}

To simulate the intermittent fluctuations around unstable steady states observed in the experimental recordings (see Fig~\ref{fig:figS3}), we used the Morris-Lecar model (eq\,(8) of the main manuscript) with added external current spikes in the $V$-equation. Namely, we replaced the $V$-equation from eq\,(8) by
\begin{equation}\label{eq:MLspikes}
\begin{split}
C\mathrm{d}V &= \bigg(- g_\mathrm{L}(V-V_\mathrm{L})-g_{Ca}m_{\infty}(V)(V-V_{Ca})-g_{K}w(V-V_{K})\ldots\\ &\ldots+g_c(V_{\mathrm{h}}-V)+\sum_i I_i(t)\bigg)\mathrm{d}t+\sigma_W \mathrm{d}W
\end{split}
\end{equation}
where $I_i(t)=I_p\delta(t-t_i)$ is a current spike modelled by a Dirac delta input, which we approximated by two Heaviside functions. Namely: $\delta(t-t_i)\approx\text{Heav}(t-t_i)\ast\text{Heav}(t_i+t_\mathrm{dur}-t)$, with the duration of the spike chosen to be $t_\mathrm{dur}=10$ms, and its amplitude being $I_p=50$pA. We also added noise in the form of a Wiener process $W$ of amplitude $\sigma_W=0.5$. Other parameters are as in \emph{Materials and Methods} of the main manuscript. For the loss of contact simulation (Fig\,DB in S1 Text) we used a piecewise-constant $g_c$ dropping from $20$ to $5$ along the unstable branch of equilibria (other parameters as in \emph{Materials and Methods} of the main manuscript) and we also added noise using a Wiener process $W$ as in the previous case, of amplitude $\sigma_W=0.1$.

\begin{figure*}[!b]
\centering
\includegraphics[width=1\columnwidth]{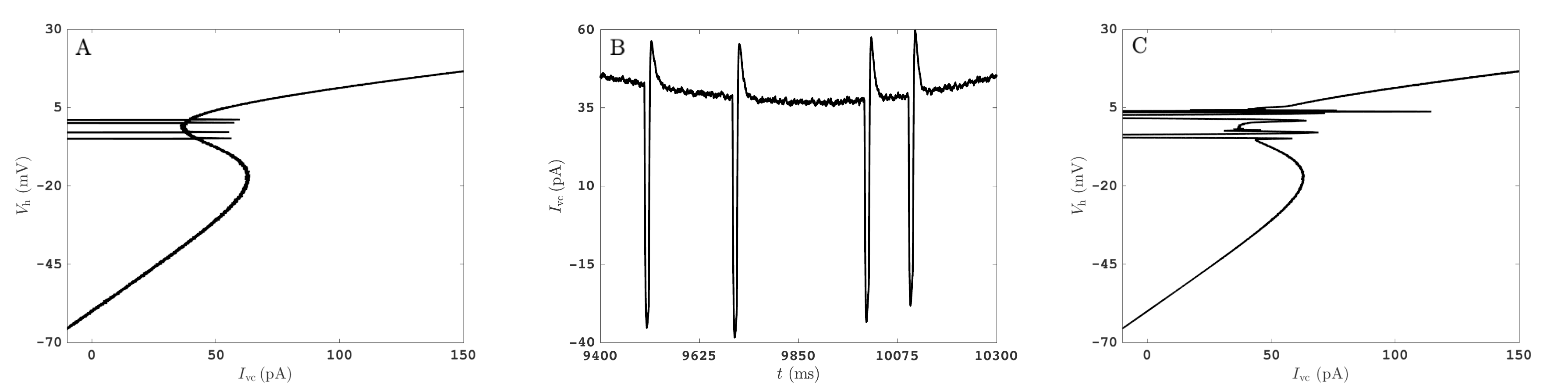}
\caption{Reproducing the experimental intermittent fluctuations with the ML model. A-B: Simulation of the ML model with current spikes artificially added to the voltage equation, viewed in the $(\Ivc,\Vh)$-plane in panel A and with the time series of $\Ivc$ in panel B. C: Simulation of the ML model with artificial loss of contact through a drop of gain $\gc$, viewed in the $(\Ivc,\Vh)$-plane.}
\label{fig:figS4}
\end{figure*}   
We hypothesise that these intermittent fluctuations are currents (possibly mediated by voltage-gated sodium and/or calcium channels) generated in distal neuronal sites (e.g., dendrites, axons) due to non-ideal voltage clamp (i.e., space-clamp error due to failure to achieve isopotential across the entire neuronal cell). 

This can be understood by first comparing I-V relationships of voltage-gated channels of neurons in slice preparations versus isolated neurons, and subsequently inferring what occurs to isolated voltage-gated channels that are poorly controlled due to space-clamp error. Neurons in slice preparations have lower input resistance and higher membrane capacitance when compared to an isolated neuron~\cite{kay1986}. We observed the same effect with our entorhinal cortical neurons, which in slice preparations had $R \in \left]150,350\right[ M\Omega$ and $C \approx 100$ pF, while in a previous study (unpublished) the isolated case had $R\in \left]500,1000\right[ M\Omega$ and $C\in \left]10,30\right[$\,pF. Thus, in slice preparations the typically observed voltage range in which the I-V relationship has a negative slope (e.g., for sodium channels) is quite large~\cite{sherman1999}. In ideal conditions (i.e., when voltage-gated channels are more or less clustered within a neuronal site, for instance, close to the soma), space-clamp error is minimised. In this case, when the membrane voltage is slowly increased then individual negative peak currents of voltage-gated channels are successively activated (i.e., a kind of averaging effect occurs due to the clustering of channels), which give rise to a wide voltage range in which the I-V relationship has a negative slope. In other words, the current peaks are distributed across the voltage range of the I-V relationship~\cite{sherman1999}. 

Noteworthy, the fast sodium current peaks occur during this negative slope phase because the voltage dependence of the transient sodium channel activation (which mostly mediate these peaks) is similar to the activation of the persistent sodium channels. The poorly voltage-clamped transient sodium channels mediate the fast negative currents, while the transient ones mediate a slow negative current, both of which appear in the same voltage range~\cite{fleidervish1996}. In contrast, isolated neurons typically lose most of their thin dendritic branches (i.e., abbreviated morphology), hence voltage-gated channels (e.g., sodium current) display only one negative peak threshold current under a step depolarisation~\cite{kay1986,golding1999}. From these observations, it is reasonable to assume that in non-ideal conditions (i.e., space-clamp error) there are isolated channels furthest from the cell soma or electrode contact (e.g. distal dendrites with thin branches, or dendritic spines with increased resistance of the cytoplasm, or distal sodium channels)~\cite{dumenieu2017, petersen2015}. These isolated channels have I-V relationships with negative slope but with a shorter voltage range and distinct activation threshold. Thus, under a VC protocol the mismatch between the hold voltage ($\Vh$) and the real-membrane voltage ($V$) of distal voltage-gated channels can be very large, resulting in the observed intermittent fluctuations. Additionally to the above mentioned factors, the delay of the feedback (associated with the amplifier, which is in the order tens of nanoseconds) is likely to play a role in these observed intermittent fluctuations. However, from a theoretical standpoint, a fixed delay can induce a Hopf bifurcation, and therefore, robust oscillations. So instead, we envisage a varying delay, due to possibly latent processes, could create this effect. Of course, to really address this point, further experiments are required.

To simulate the above hypothesis we added additional instantaneous currents (i.e. to model the effect of non-ideally clamped channels as external forcing that correspond to non-homogeneous terms); see Section \textbf{Computation models} of the main manuscript for details, in particular eq.\,(10). The results are depicted in Fig~\ref{fig:figS4} in the $(I,V)$-plane (panel A) and as a corresponding time series $\Ivc(t)$ (panel B). Both, time profile and $(\Ivc(t),\Vh(t))$-curve closely resemble the fluctuations in the experimental data in Fig\,CA in S1 Text. Although, these simulations may explain the experimental observations, we do not exclude other possible mechanisms (or a mix of mechanisms). To this end, note that the standard VC protocol was designed to stabilise the membrane voltage $V$ at some static reference signal $V_h$, for a given, large enough, control gain $g_c$. However, our VC protocol involves slowly varying $V_h$, which causes the neuronal membrane $V$ to quasi-statically stabilise (drift) without ever reaching the true steady state. Along the unstable branch, the repulsion (associated with unstable eigenvalues) may not necessarily be constant and thus the degree of effectiveness of the control signal $g_c (V_h- V)$ may vary. Noteworthy, this control signal is effectively the proportional control term in the proportional–integral–derivative (PID) control formalism. Thus, it is also possible the proportional control does not achieve a full control of more complex physiological processes in which case the control may only be intermittent. Other possibilities may include temporary loss of contact between the electrode and the neuronal membrane. To elucidate these points, we simulated the Morris-Levar equation (see \emph{Material and Methods} in main manuscript) but with variations in $g_c$. Specifically, we considered a piecewise-constant $g_c$ corresponding to decreasing its default value (thus modelling loss of control) along the unstable branch of equilibrium; see Fig\,DC in S1 Text for a simulation. While this scenario bears less resemblance to the experimental observation, it can either be used to rule out this possibility or to consider new experiments to further clarify these observations.

\section{Experiments estimating hysteresis effects and blocking of voltage-gated sodium and calcium ion channels}
\label{sec:hysteresis}
The voltage-clamp recordings of the responses to the slow ramps of command voltage (Fig\,EA in S1 Text) were made to obtain the current-voltage relationships (I-V relationships).  With the potassium gluconate-based pipette solution being used, the I-V relationship obtained by applying the ascending ramp was not the same as the one obtained by applying the descending ramp (Fig\,EB in S1 Text). The ascending ramp produced an S-shaped I-V relationship, which had the negative slope in the voltage range of $-50$ to $-30$\,mV (Fig\,EB in S1 Text, CTRL, green trace). The descending ramp of the command voltage produced an I-V curve with a positive slope throughout the physiological voltage range. 
\begin{figure*}[!h]
\centering
\includegraphics[width=0.6\columnwidth]{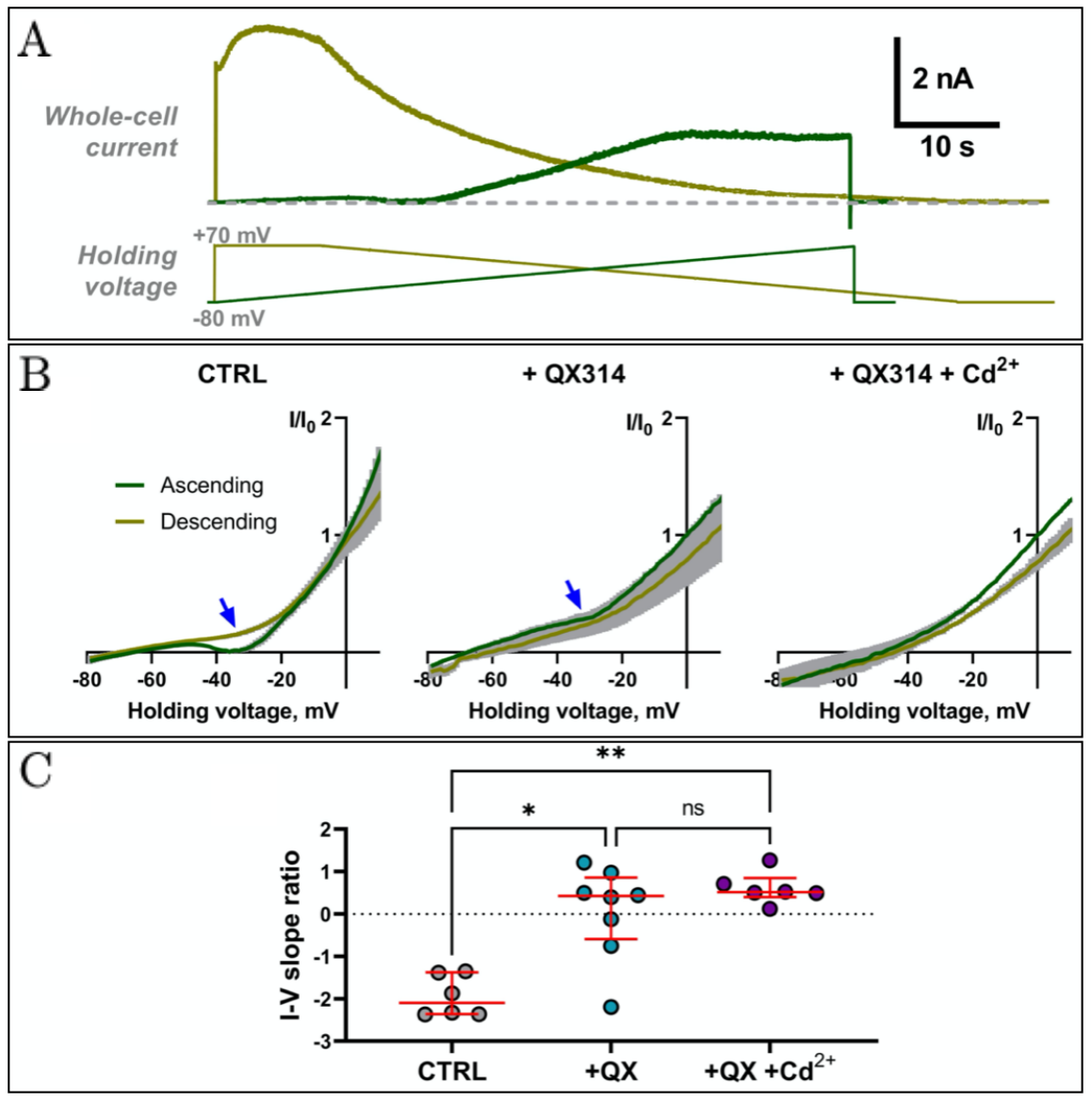}
\caption{A: The representative voltage-clamp recordings of the responses to the slow ramps of the command potential. Top traces represent the whole-cell current, bottom trace represent the command potential protocol. Dark green lines correspond to the ascending ramp; light green lines correspond to the descending ramp. B: Pairs of averaged ascending and descending I-V relationships, obtained in control conditions (left panel), with QX314 in the pipette solution (middle panel), and with both QX314 in the pipette solution and Cd$^{2+}$ in the ACSF (right panel). Gray bars represent the standard error. The I-V relationships were normalized to the value of current at 0 mV. C: The comparison of I-V slope ratios in three experimental conditions (Kruskal-Wallis ANOVA, P$<0.001$, followed by Dunn't post hoc test; $^*$ - P$<0.05$, $^{**}$ - P$<0.01$)}
\label{fig:figS5}
\end{figure*}

We hypothesized that the mismatch between the ascending and descending I-V relationships can be attributed to the activity of the slowly inactivating sodium and calcium ion channels. These slowly inactivating sodium and calcium channels are active at the ascending ramp, whereas high depolarization at the descending ramp inactivates them such that they are ineffective and thus do not contribute to the S-shaped I-V relationship. It has previously been reported that persistent sodium channels are present in deep entorhinal cortex neurons, and that current through these channels can be detected even when the rate of membrane depolarization is very low~\cite{agrawal2001}. A prepulse to depolarized voltages inactivates persistent sodium channels of the neocortical neurons~\cite{fleidervish1996}. T-Type calcium channels are also known to produce the ``window'' currents due to the overlap of the activation and steady-state inactivation curves~\cite{amarillo2014}. In order to check this hypothesis we used QX314 (6 mM) and Cd$^{2+}$ (50 {\textmu}M), which are the blockers of the voltage-gated sodium and calcium ion channels, respectively. To quantify the effects of the applied substances on the mismatch between the ascending and descending I-V relationships we introduced the I-V slope ratio (Fig\,EC in S1 Text). It was calculated as the ratio of the minimal slope of the ascending I-V relationship to the slope of the descending I-V relationship at -35 mV. In control conditions, the I-V slope ratio was negative due to a strong mismatch between the I-V relationships (Fig\,EB and EC in S1 Text, CTRL). When the pipette solution containing QX314 was used, the observed mismatch between the I-V relationships was much less pronounced (Fig\,EB in S1 Text, middle panel), resulting in the significantly higher I-V slope ratio. With QX314 in the pipette solution and Cd$^{2+}$ in the ACSF, both the ascending and descending I-V relationships were monotonically increasing functions (Fig\,EB in S1 Text, right panel), resulting in positive values of the I-V slope ratios. The results obtained suggest that the negative slope of the ascending I-V relationship is mainly mediated by persistent sodium channels and to a lesser extent by voltage-gated calcium channels. \rev{To avoid inactivation of these channels by high depolarization all voltage-clamp runs in the main manuscript increase the hold voltage $\Vh$, and all current-clamp runs increase the hold current $\Ih$.}

In the model the channel properties depend sensitively on the maximal conductances (e.g., $g_\mathrm{Ca} m_\infty$ for calcium in eq\,(8) of the main manuscript). A small change in these parameters in eq.\,(8) removes the two fold points from the steady-state curve (in a so-called cusp bifurcation), resulting in a change from \rev{class}-I to \rev{class}-II behavior.
\newpage
\section{Equilibrium analysis of the general conductance-based model for the $\mathbf{V}$-clamped protocol}
For a fixed hold voltage $\Vh$ a general conductance-based model, written using indices $j$ for $\nc$ channels and rates $r_j(V)=1/\tau_j(V)$, reads (see main text, equation (1)):
\begin{align}\label{eq:VCSRfast1}
C\dot{V} &=-\sum_{j=1}^\nc I_j(x_j,V)+\gc(\Vh-V),&
\dot{x}_j &=(x_{\infty,j}(V)-x_j)r_j(V).
\end{align}
This section states a sufficient criterion for the uniqueness and linear stability of equilibria for system~\eqref{eq:VCSRfast1}. Equilibria are at
$x_{\mathrm{eq},j}=x_{\infty,j}(\Veq)$, where $\Veq$ satisfies
\begin{align}\label{veq:fixedpoint}
  \Veq&=\Vh-\frac{1}{\gc}\sum_{j=1}^\nc I_j(x_{\infty,j}(\Veq),\Veq)=:\afix(\Veq).
\end{align}
We denote the $\nc\times 1$ vectors ($j=1,\ldots,\nc$)
\begin{align*}
  I_{\infty,j}&=I_j(x_{\infty,j}(\Veq),\Veq),
  &  x_{\infty,j}'&=x_{\infty,j}'(\Veq),\\
  I_{x,j}&=\partial_1I_j(x_{\infty,j}(\Veq),\Veq),
  &  r_{\mathrm{eq},j}&=r_j(\Veq),\\
  I_{V,j}&=\partial_2I_j(x_{\infty,j}(\Veq),\Veq),
  &  r_{\mathrm{eq},j}'&=r_j'(\Veq),\\
&& \tau_{\mathrm{eq},j}&=\tau_j(\Veq).
\end{align*}
The derivative of $\afix(V)$ in $V=\Veq$ is (using the notation $\bm{1}=(1,\ldots,1)^\tran$)
\begin{align*}
  \afix(\Veq)'&=-\frac{1}{\gc}[I_x^\tran x_\infty'+I_V^\tran\bm{1}]=:\muc,
\end{align*}
which is dimensionless.  Consequently, a sufficient condition for the
uniqueness of the fixed point $\Veq$ is $|\muc|<1$, or, which is guaranteed by the estimate
\begin{align*}
  \mc:=\frac{1}{\gc}[\|I_x.x_\infty'\|_1+\|I_V\|_1]<1,
\end{align*}
where $(.)$ means element-wise multiplication (so, $v.w=\diag(v)w$), and $\|v\|_1:=|v_1|+\ldots+|v_{n_\mathrm{c}}|$ is the $1$-norm of a vector $v$. The
condition $\mc<1$ is also sufficient for linear stability of
$(\Veq,x_{\mathrm{eq},(\cdot)})$: the Jacobian of
\eqref{eq:VCSRfast1} in $(\Veq,x_{\mathrm{eq},(\cdot)})$ is
\begin{align*}
  J_\mathrm{eq}:=
  \begin{bmatrix}
    -(\gc+I_V^\tran\bm{1})/C,&-I_x^\tran/C\\[0.5ex]
    r_\mathrm{eq}.x_\infty',&-\diag(r_\mathrm{eq})
  \end{bmatrix},
\end{align*}
(note that the matrix entries are in
row order $1\times1$, $1\times \nc$, $\nc\times 1$ and
$\nc\times \nc$, and that the term
$\diag(r_\mathrm{eq}')x_{\infty}(\Veq)$ cancels in the bottom
left entry).  By rescaling the $x_j$ to $x_j/(x_{\infty,j}'(1+\delta))$
with an arbitrary $\delta>0$ we can make the lower $\nc$ rows diagonally dominant. This gives then in the upper right entry $-(I_x.x_\infty')^\tran(1+\delta)/C$:
\begin{align*}
  J_\mathrm{scal}:=
  \begin{bmatrix}
    -(\gc+I_V^\tran\bm{1})/C,&-\frac{1+\delta}{C}(I_x.x_\infty')^\tran\\[0.5ex]
    r_\mathrm{eq}/(1+\delta),&-\diag(r_\mathrm{eq})
  \end{bmatrix},
\end{align*}
Typically, the derivative of $I_j$ with respect to $V$ is non-negative (so, $I_{V,j}\geq0$).
Hence, the matrix satisfies Gershgorin's theorem for some $\delta>0$ if
\begin{align*}
  \gc+\sum_{j=1}^\nc I_{V,j}&>\sum_{j=1}^\nc|I_{x,j}x_{\infty,j}'|\mbox{,\quad or\quad}
  \gc+\|I_V\|_1>\|I_x.x_\infty'\|_1.
\end{align*}
This is equivalent to $\mc<1$. The fixed-point relation \eqref{veq:fixedpoint} also ensures that
\begin{align*}
  \Vh(1-\mc)\leq \Veq\leq \frac{\Vh}{1-\mc},
\end{align*}
which gives a bound on the relationship between $\Vh$ and $\Veq$, that becomes sharper for larger $\gc$. Hence, for conductance-based models with clamped voltage $\Vh$ the difference between the clamp voltage $\Vh$ and the unique stable equilibrium potential across the membrane is of order $\mc$ (so gets small for large gain $\gc$).

\section{Conditions for validity of estimate for systematic bias due to dynamic ramp of voltage} \label{sec:projection}
Estimate (5) in the main text provides an approximation to first order in $\eps$ for the bias between steady-state current, $\Ieq(\Vh)$, for fixed hold voltage $\Vh$, and the actually measured current $\Ivc(t)$ when the hold voltage is changed slowly with $\dotVh=\eps\Delta_V$: 
\begin{align}\label{eq:SM:close}
    \Ieq(\Vh(t))-\Ivc(t)\approx\Ieq'(\Vh)\dotVh/\alpha.
\end{align} 
In this expression $\alpha$ is the decay rate toward the equilibrium $\Ieq(\Vh)$ at fixed $\Vh$ and $\Ieq'(\Vh)$ is the slope of the equilibrium (steady-state) curve $\Ieq(\Vh)$ we are interested in. In practice, we estimate 
\begin{align}\label{proj:SM:ieqalpha:est}
    \Ieq'(\Vh(t))&\approx\Ivc'(\Vh(t)),&
    \alpha^{-1}&\approx\tau_{1/2}/\ln2,
\end{align} where $\Ivc(\Vh)$ is the measured $(I,V)$-curve during voltage clamp runs (VC) and $\tau_{1/2}$ is the time of decay to half of the transients during large fluctuations of measured $\Ivc$ during VC runs (see Section~\ref{sec:fluctuations} and Fig~\ref{fig:figS4}). The estimates \eqref{proj:SM:ieqalpha:est} are themselves accurate to first order in $\eps$, such that their bias is only a second-order effect in \eqref{eq:SM:close} if $\alpha>0$.
Pretending $\Ivc(t)$ was governed by a scalar first-order ordinary differential equation $\dot{I}_\mathrm{vc}=f_{IV}(\Ivc,\Vh)$, bias estimate \eqref{eq:SM:close}  would follow from implicit differentiation with respect to time and $\Ivc$ , where $\alpha=-\partial_1f_{IV}(\Ieq(\Vh),\Vh)>0$.

Justifying \eqref{eq:SM:close} based on a more accurate model for the VC experiment such as a higher-dimensional conductance-based model requires that two conditions are met:
\begin{compactitem}
    \item presence of a dominant eigenvalue with \emph{spectral gap} in the linearization near the steady-state curve, and
    \item \emph{observability} of fluctuations with the slowest rate of decay in the measured current.
\end{compactitem}
Examples of such models are equations (1), (2) in the main text; see also the Morris-Lecar model used for Fig\,3 of the main manuscript and Fig~\ref{fig:figS4}, given in \emph{Materials and Methods} in the main manuscript.
More precisely, let us collect  the entire vector of conductance-based model variables into $y=(V,x_j)_{j=1,\ldots,n_\mathrm{c}}$, such that the conductance-based model (1),\,(2) for voltage clamp with a slow ramp of hold voltage $\Vh$  from the main text is of the general form
\begin{align}\label{gen:SM:slowfast}
\dot y&=f(y,\Vh),&\dotVh&=\eps\Delta_V.
\end{align}
For each fixed hold voltage $\Vh$ (so, at $\eps=0$) the voltage clamp stabilizes an equilibrium $\yeq(\Vh)$, such that $0=f(\yeq(\Vh),\Vh)$. 

\paragraph{Spectral gap} The first condition for validity of \eqref{eq:SM:close} is that among the eigenvalues of the matrix $J(\Vh):=\partial_1f(\yeq(\Vh),\Vh)$ (which all have negative part), there is one \emph{dominant} eigenvalue $-\alpha(\Vh)$ of $J(\Vh)$ with largest real part. This dominance is expressed as a spectral gap condition for $J(\Vh)$, which means that all other eigenvalues must have real part less than some bound $-\beta(\Vh)<-\alpha(\Vh)<0$ with the condition
\begin{align}\label{SM:specgap}
    \beta^{-1}(\Vh)&\ll\alpha^{-1}(\Vh)\sim \mbox{\ order 1.}
\end{align}
The current flowing through the membrane is the scalar output $\Ivc(y)=\sum_jI_j(x_j,V)\in\mathbb{R}$, depending on the state variables of the conductance-based model, which we measure. Near the equilibrium $\yeq(\Vh)$ this output is to first order of the form
\begin{align*}
    \Ivc(y)\approx \Ivc(\yeq(\Vh))+\partial \Ivc(\yeq(\Vh))[y-\yeq(\Vh)]\mbox{,\quad where $\partial \Ivc(\yeq(\Vh))=:c(\Vh)^\tran\in\mathbb{R}^{1\times(n_\mathrm{c}+1)}$ is a row vector.}
\end{align*}
\paragraph{Observability} The second condition for validity of \eqref{eq:SM:close} is that the right eigenvector $\yr(\Vh)$ for the dominant eigenvalue $-\alpha(\Vh)$ has a non-zero (more precisely, non-small) projection in variations of this measured output:
\begin{align}\label{SM:outputproj}
    \partial \Ivc(\yeq(\Vh))\yr(\Vh)=c(\Vh)^\tran\yr(\Vh)\neq0\mbox{.}
\end{align}
If the dominant direction $\yr$ is observable according to \eqref{SM:outputproj}, then we may normalize the eigenvector $\yr$ such that $c(\Vh)^\tran\yr(\Vh)=1$ for all $\Vh$. We denote the left eigenvector of $J(\Vh)$ for the dominant eigenvalue $-\alpha(\Vh)$ by $\yl(\Vh)$, normalized such that $\yl(\Vh)^\tran\yr(\Vh)=1$. If $J(\Vh)$ satisfies the spectral gap condition \eqref{SM:specgap}, its inverse is  $J(\Vh)^{-1}\approx -\alpha(\Vh)\yr(\Vh)\yr(\Vh)^\tran$ (ignoring small terms of order $\beta^{-1}$), thus, it is close to the rank-$1$ matrix $-\alpha\yr\yl^\tran$. As $\yeq (\Vh)$ is the steady-state curve of \eqref{gen:SM:slowfast} (so, $0=f(\yeq(\Vh),\Vh)$), we have (abbreviating $\partial_2f_\mathrm{eq}(\Vh):=\partial_2f(\yeq(\Vh),\Vh)$ and dropping argument $\Vh$ from $\alpha$, $c$, $\yr$, $\yl$, $J$, and $\partial_2f_\mathrm{eq}$) 
\begin{align*}
    \yeq'(\Vh)&=-J^{-1}\partial_2f_\mathrm{eq}\approx -\alpha\yr\yl^\tran\partial_2f_\mathrm{eq}\mbox{, \quad and, hence,}\\
    \Ieq'(\Vh)&=\partial\Ivc(\yeq(\Vh))\yeq'(\Vh)=c^\tran\yeq'(\Vh)\approx -\alpha\yl^\tran\partial_2f_\mathrm{eq}\mbox{,\quad implying}\\
    \yeq'(\Vh)&\approx\yr\Ieq'(\Vh)
\end{align*}
when neglecting terms of order $\beta^{-1}$. Expansion of \eqref{gen:SM:slowfast} near the steady-state curve $(\yeq(\Vh),\Vh)$ implies that solutions $y(t)$ of \eqref{gen:SM:slowfast} satisfy (after transient have decayed, dropping the argument $t$ from $\Vh$) 
\begin{align*}
    y(t)-\yeq(\Vh(t))\approx J^{-1}\yeq'(\Vh)\dotVh\approx-\alpha\yr\yl^\tran\yr \Ieq'(\Vh)\dotVh=-\alpha\yr \Ieq'(\Vh)\dotVh,
\end{align*}
ignoring terms of order $\eps^2$ and $\eps\beta^{-1}$. Inserting this expansion into the expansion for the output $\Ivc(y(t))$ gives the desired first-order bias estimate,
\begin{align}\label{SM:outputapprox}
    \Ivc(y(t))-\Ieq(\Vh(t))\approx \partial\Ivc(\yeq(\Vh))[y(t)-\yeq(\Vh(t))]\approx -\alpha c^\tran\yr\Ieq'(\Vh)\dotVh=-\alpha \Ieq'(\Vh)\dotVh.
\end{align}
Finally, we observe that the measured current $\Ivc(t)$ when $\Vh$ changes dynamically is not $-\sum_jI_j(x_j,V)$ but (as measured at the electrode) equal to $\gc(\Vh-V)$. However, $\Ivc(t)=\gc(\Vh(t)-V(t))=\sum_jI_j(x_j,V)+C\dot V(t)$, where $\dot V$ is a component of $\dot y\approx f(y,\Vh)\approx J[y(t)-\yeq(\Vh(t))]$, which equals $\yeq'(\Vh)\dotVh$ and is of order $\eps$ after transients have decayed. Thus, the correction due to this difference in \eqref{SM:outputapprox} is only another $O(\eps)$ term in $\Ieq'(\Vh)$, making it a second-order effect overall.
\clearpage
\section{Numerical bifurcation diagrams for in-silico models}
\label{sec:sec:numbif}
\begin{figure}[ht]
  \centering
  \includegraphics[width=0.9\textwidth]{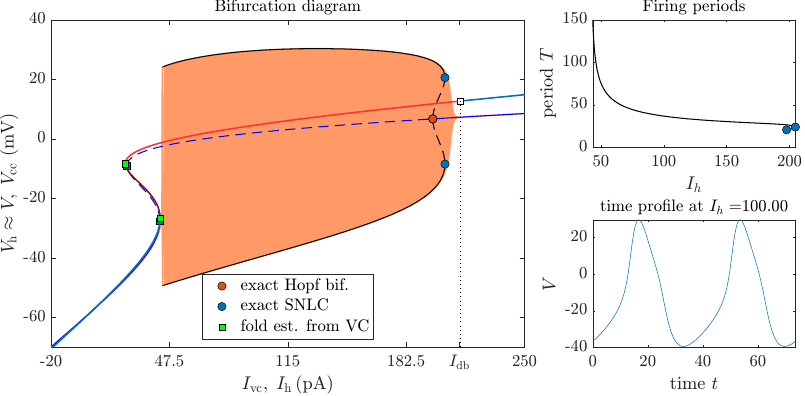}
  \caption[Numerical bifurcation diagram for \rev{class}-I Morris-Lecar model]{Numerical bifurcation diagram overlaid for Fig\,3A from the main manuscript, showing a \rev{class}-I Morris-Lecar neuron model~\cite{morris1981} as example of excitatory cell. Black curves show $\max_tV(t)$ and $\min_tV(t)$ for periodic firing oscillations at each fixed $\Ih$ (solid for dynamically stable, dashed for dynamically unstable). Equations and parameter values are given in \emph{Material and Methods} of the main manuscript. The diagrams on the right show the dependence of the period of the periodic orbits on the parameter and a typical periodic time profile for $\Ih=100$\,pA.}
  \label{fig:figSI6}
\end{figure}
\begin{figure}[ht]
  \centering
  \includegraphics[width=0.9\textwidth]{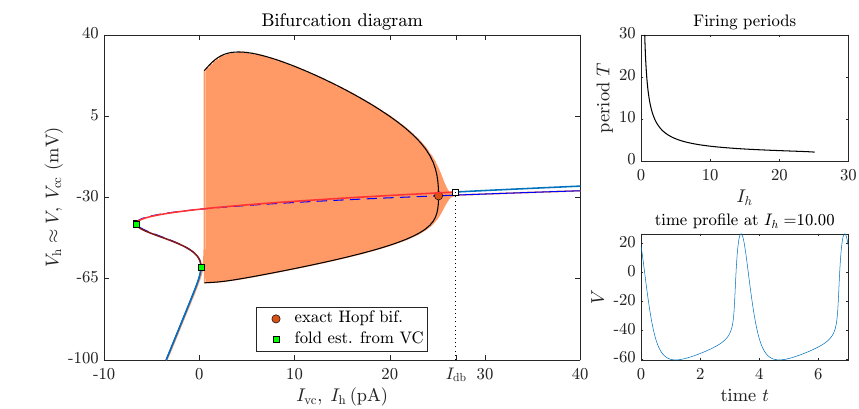}
  \caption[Numerical bifurcation diagram for Wang-Buzaki model]{Numerical bifurcation diagram overlaid for Fig 3B from the main manuscript, showing a Wang-Buzsaki neuron model \cite{wang1996} as example of inhibitory interneuron. Black curves show $\max_tV(t)$ and $\min_tV(t)$ for periodic firing oscillations at each fixed $\Ih$ (solid for dynamically stable, dashed for dynamically unstable). Equations and parameter values are given in \emph{Material and Methods} of the main manuscript.  The diagrams on the right show the dependence of the period of the periodic orbits on the parameter and a typical periodic time profile for $\Ih=10$\,pA.}
  \label{fig:figSI7}
\end{figure}
\begin{figure}[ht]
  \centering
  \includegraphics[width=0.9\textwidth]{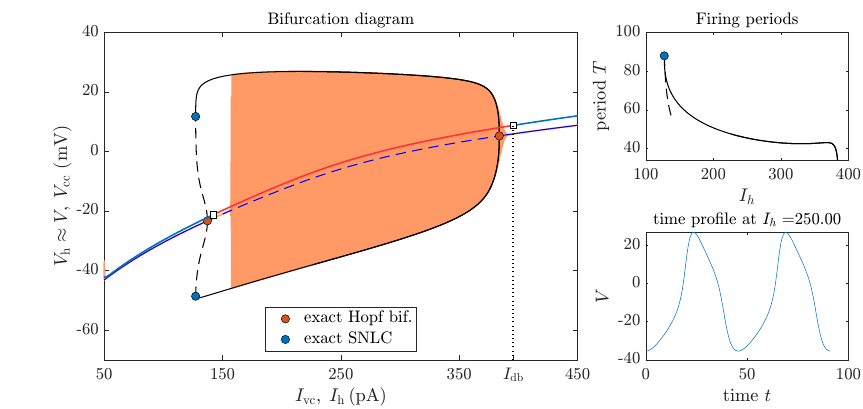}
  \caption[Numerical bifurcation diagram for \rev{class}=II Morris-Lecar model]{Numerical bifurcation diagram overlaid for Fig 4 from the main manuscript, showing a \rev{class}-II Morris-Lecar neuron model \cite{morris1981} as example of \rev{class}-II neuron. Black curves show $\max_tV(t)$ and $\min_tV(t)$ for periodic firing oscillations at each fixed $\Ih$ (solid for dynamically stable, dashed for dynamically unstable). Equations and parameter values are given in \emph{Material and Methods} of the main manuscript.  The diagrams on the right show the dependence of the period of the periodic orbits on the parameter and a typical periodic time profile for $\Ih=250$\,pA.}
  \label{fig:figSI8}
\end{figure}
The diagrams in Figs~F, G and H in S1 Text contain copies of the graphs of VC and CC protocols applied to the Morris-Lecar model \cite{morris1981} and the Wang-Buzaki model \cite{wang1996}, as shown in Figs\,3 and 4 of the main manuscript. All equations and parameter values are given in \emph{Materials and Methods} of the main manuscript. Figs~F, G and H in S1 Text also contain overlaid numerical bifurcation diagrams obtained using \textsc{coco} \cite{DS13}. The code reproducing the figures is available at \url{https://github.com/serafimrodrigues/Amakhin-etal-2025}. The repository includes a snapshot of \textsc{coco} (revision 3328) as downloaded from \url{https://sourceforge.net/projects/cocotools}.

Near Hopf bifurcations at $I_\mathrm{db}$ in Fig~\ref{fig:figSI6} and at $\Ih\approx 135$\,pA in Fig~\ref{fig:figSI8} we observe a small region of bistability. This region shrinks for stronger time scale separation (parameter $\phi$ in the Morris-Lecar model).
\clearpage
\section{Zoom near $I_\mathrm{db}$ for CC protocol of cell 5}
\label{sec:zoom_Idb}
\begin{figure}[ht]
  \centering
  \includegraphics[width=0.6\textwidth]{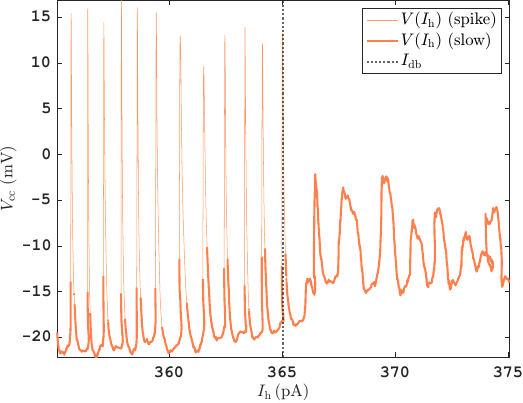}
  \caption[Zoom of CC run for cell 5 near $I_\mathrm{db}$]{Zoom of CC run for cell 5 near the estimate for $I_\mathrm{db}\approx365$\,pA. Parts of the recordings where smoothed time profiles of $(\Vcc'(t),\Vcc''(t))$ have a norm smaller than a threshold are highlighted by a thicker line and labelled as slow part of the firing oscillation or near equilibrium.}
  \label{fig:figSI9}
\end{figure}
Fig~\ref{fig:figSI9} shows a zoom of the CC run data for cell 5 of Fig\,1B of the main manuscript. 
\paragraph{Highlighting of slow part of voltage profile in CC run}
The criterion for labelling part of the time series $\Vcc$ of a CC run as \emph{slow} is if the time profile $(\Vcc',\Vcc'')$ has a norm less than some threshold $\theta_\mathrm{th}$ at time $t_i$ for sampling times $t_i$. More precisely, let $(t_i,V_i)$ be the voltages $V_i$ sampled at times $t_i$ for $i=1,\ldots,n_\mathrm{s}$. We define the operations
\begin{align*}
    t^{(k)}&=\mbox{\ sampling times $t$, refined by $k$ uniformly spaced }\\[-0.2ex]
    &\phantom{=}\mbox{\quad intermediate times between $t_{i-1}$ and $t_i$ for each $i=2,\ldots,n_\mathrm{s}$,}\\
    V^{(k)}&=\mbox{\ linear interpolation of $(t_i,V_i)$ at $t^{(k)}$,}\\
  \operatorname{sm}_{w}x&=\mathtt{smoothdata(}x\mathtt{,"gaussian",}w\mathtt{, "SamplePoints",}t^{(k)}\mathtt{)}\\
  \operatorname{sp}x&=\mathtt{interp1(}t^{(k)},x\mathtt{,"spline","pp")}\\
  \operatorname{ev}f&=f(t^{(k)})
  \end{align*}
(with the Matlab commands \texttt{smoothdata} for smoothing, \texttt{interp1} for cubic spline interpolation), where we note that $\operatorname{sp}$ is applied to a profile (vector) sampled at $t^{(k)}$, resulting in a differentiable function, while $\operatorname{ev}$ applies a function on the sample points $t^{(k)}$ resulting in a vector sampled on $t^{(k)}$. With these operations we construct a smoothed differentiation for a profile $x$ on $t^{(k)}$ as
\begin{align*}
  x'&=\operatorname{ev}\circ \frac{\mathrm{d}}{\mathrm{d} t}\circ \operatorname{sp}\circ\operatorname{sm}_w^2 x.
\end{align*}
With this definition $\Vcc$ and $\Ih$ at all times $t^{(k)}_i$ for which
\begin{align*}
  \left({V^{(k)}}''\right)^2\Delta_t^2+\left({V^{(k)}}'\right)^2\Delta_t&<\theta_\mathrm{th}^2
\end{align*}
as slow (thick line) in Figs~\ref{fig:figSI9}, 1B and 2 of the main manuscript. Parameter values were $k=5$, $w=0.01$\,s, $\theta_\mathrm{th}=5$ for Figs\,1B, 2, A and I in S1 Text. The refining by $k$ additional points in for the sampling $t_i$ enables the application of the operation $\operatorname{sm}_w$ with a window size close to the sampling time.
\end{document}